%% file: main.tex
\documentclass[10pt, conference, compsocconf]{IEEEtran}
\usepackage{graphicx}
\usepackage{epsfig}
\usepackage{lipsum,multicol}
\usepackage{amsmath}
\usepackage{xspace}
\usepackage{comment}
\usepackage{flushend}
\usepackage{array}
\usepackage{threeparttable}
\usepackage[ruled,vlined,linesnumbered]{algorithm2e}
\usepackage{hyperref}

\DeclareMathOperator*{\argmin}{argmin}
\newcolumntype{P}[1]{>{\centering\arraybackslash}p{#1}}
\newcolumntype{M}[1]{>{\centering\arraybackslash}m{#1}}
\newcommand{\ie}{\emph{i.e.,}\xspace}
\newcommand{\eg}{\emph{e.g.,}\xspace}

\SetKwProg{Fn}{Function}{}{}

\SetCommentSty{mycommfont}

\newcommand{\cv}{{\bf c}}
\newcommand{\mv}{{\bf m}}
\newcommand{\xv}{{\bf x}}
\newcommand{\onev}{{\boldsymbol 1}}

\newcommand{\name}{{I-BOT}\xspace}

\newcommand{\nameI}{{I-BOT-I$^2$}\xspace}
\newcommand{\nameP}{{I-BOT-PI}\xspace}

\newtheorem{lemma}{Lemma}
\usepackage{mathtools}
\DeclarePairedDelimiter\ceil{\lceil}{\rceil}
\DeclarePairedDelimiter\floor{\lfloor}{\rfloor}
\newcommand{\ind}{\boldsymbol{1}}

\title{{I-BOT: Interference-Based Orchestration of Tasks for Dynamic Unmanaged Edge Computing 
}}

\AtBeginDocument{%
  \providecommand\BibTeX{{%
    \normalfont B\kern-0.5em{\scshape i\kern-0.25em b}\kern-0.8em\TeX}}}

\usepackage{subcaption}

\usepackage{dblfloatfix}

\begin{document}
\pagestyle{plain}
\author{Shikhar Suryavansh, Chandan Bothra, Kwang Taik Kim, Mung Chiang, Chunyi Peng, Saurabh Bagchi \\
Purdue University \\
\{ssuryav, cbothra, kimkt, chiang, chunyi,  sbagchi\}@purdue.edu
}
% \affiliation{\large Purdue University}
% \email{e-mail: {ssuryav, kimkt, cbothra, chiang, chunyi,  sbagchi}@purdue.edu  }
\vspace{15pt}

\maketitle

% \newpage

\input{sec_abstract}

\input{sec_introduction}

\input{sec_motivation_challenges}

\input{sec_system_overview}

\input{sec_design}

\input{sec_evaluation}

% \newpage{\phantom{blah}}
% \newpage{\phantom{blah}}

\input{sec_discussion}

\input{sec_related_work}

\input{sec_conclusion}

% \input{sec_acknowledgments}

% \newpage{\phantom{blah}}
% \newpage

\bibliographystyle{acm}
\bibliography{biblio}

% \balancecolumns
\end{document}

%% file: sec_abstract.tex
\vspace{4 cm} 
\begin{abstract}

In recent years, edge computing has become a popular choice for latency-sensitive applications like facial recognition and augmented reality because it is closer to the end users compared to the cloud. Although infrastructure providers are working toward creating managed edge networks, personal devices such as laptops, desktops, and tablets, which are widely available and are underutilized, can also be used as potential edge devices. We call such devices \textit{Unmanaged Edge Devices (UEDs)}. Scheduling application tasks on such an unmanaged edge system is not straightforward because of three fundamental reasons---heterogeneity in the computational capacity of the UEDs, uncertainty in the availability of the UEDs (due to the devices leaving the system), and interference among multiple tasks sharing a UED. In this paper, we present \name, an interference-based orchestration scheme for latency-sensitive tasks on an Unmanaged Edge Platform (UEP). It minimizes the completion time of applications and is bandwidth efficient. \name brings forth three innovations. First, it profiles and predicts the interference patterns of the tasks to make scheduling decisions. Second, it uses a feedback mechanism to adjust for changes in the computational capacity of the UEDs and a prediction mechanism to handle their sporadic exits, both of which are fundamental characteristics of a UEP. Third, it accounts for input dependence of tasks in its scheduling decision (such as, two tasks requiring the same input data). To demonstrate the effectiveness of \name, we run real-world unit experiments on UEDs to collect data to drive our simulations. We then run end-to-end simulations with applications representing autonomous driving, composed of multiple tasks. We compare to two basic baselines (random and round-robin) and two state-of-the-arts, Lavea [SEC-2017] and Petrel [MSN-2018] for scheduling these applications on varying-sized UEPs. Compared to these baselines, \name significantly reduces the average service time of application tasks. This reduction is more pronounced in dynamic heterogeneous environments, which would be the case in a UEP.

\end{abstract}

%% file: sec_introduction.tex
\section{Introduction}
\label{sec:introduction}
%-------------------------------------------------------------------------------
A lot of service providers host their applications on the cloud because of the benefits of reliability, scalability and cost-effectiveness. 
% According to a recent survey~\cite{advancement_of_cloud_computing}, 90\% of the companies are hosted on the cloud. 
However, with an increase in the number of latency-sensitive applications like artificial intelligence, cloud gaming, and augmented reality, there has been a rising interest in edge computing~\cite{edge_computing}. Cloud being far from the end users is unable to support the stringent latency requirements of such applications. In the case of edge computing, computational resources are placed closer to the end users, thereby reducing latency.
Although hosting applications on the edge is attractive, edge computing adds several challenges for service providers that are distinct from cloud computing. The computational resources on the edge are heterogeneous and may not be as powerful or dependable as the cloud computing servers. Also, the available edge resources are at varying geographical distances from the users and the closest resource may not always provide the optimal service time for the applications. Therefore, when multiple users simultaneously send requests pertaining to different applications in an edge computing scenario, selecting the “best” edge device to serve the requests is non-trivial. 

We introduce the concept of {\bf ``unmanaged edge"} in this paper. Terms such as ``cloudlets"~\cite{cloudlets_1, cloudlets_2}, ``micro data centers"~\cite{mdc_1,mdc_2}, and ``fog"~\cite{fog_1,fog_2} have been used in the literature to refer to small, edge-located data centers. These cloudlets are managed by infrastructure providers such as Amazon~\cite{lambda_edge}, Cisco~\cite{cisco_edge}, and Google~\cite{google_edge}. However, with an increasing interest in executing latency-critical applications on the edge, and the personal devices such as laptops/desktops/tablets becoming more powerful than ever, there is a scope of utilizing these devices as potential edge ``servers". We call such devices {\bf ``Unmanaged Edge Devices" ($UEDs$)}. This can be thought of as moving a step closer to the end users in the user-edge-cloud continuum. 
A possible scenario where unmanaged edge can be useful is in real-time road traffic analytics using the video feed from traffic signal cameras. This involves significant video processing in real time to detect events like road accidents, traffic congestion, overspeeding, etc. Since this application is highly latency sensitive, using cloud for the processing would not suffice. In this scenario, the unmanaged edge devices available in the vicinity can be utilized.

\noindent {\bf Our Solution: \name} \\
In this paper, we present {\bf \name}, {\bf I}nterference-{\bf B}ased {\bf O}rchestration of {\bf T}asks, for unmanaged edge computing. \name optimizes the service time and the bandwidth utilization of complex applications with multiple tasks that are to be offloaded to the UEDs. We focus on task orchestration in an {\em unmanaged} edge because of the following reasons. First, managed edge devices are not yet widespread and obviously such infrastructure deployment requires significant cost and efforts to make them ubiquitous. Second, almost everyone today has powerful computing devices which are rarely utilized to their capacity. We propose using these existing underutilized resources instead of investing heavily in the infrastructure for managed edge. We verified the feasibility of unmanaged edge through a user survey (Figure~\ref{fig:survey_results}) in which $86.4\%$ of the participants indicated their willingness to participate in unmanaged edge computing (under one of four proposed incentive schemes).

% (shown in Figure~\ref{fig:user-edge-cloud-continuum})

% \begin{figure}[tb]
% \centering
% %\includegraphics[clip,trim=0cm 6cm 2.2cm 0cm,width=0.65\textwidth]{overview_diagram.pdf}
% \includegraphics[width=0.82\columnwidth]{Figures/user-edge-cloud.png}
% \vspace{-10pt}
% \caption{User-Edge-Cloud continuum}
% % SB (7/3/20): The ``Channel" component is not clear. Drop it?
% \vspace{-10pt}
% \label{fig:user-edge-cloud-continuum}
% \end{figure} 

There has been a significant amount of work~\cite{petrel,lavea,MSGA,zhang2017energy,Tseng2018} on scheduling tasks in a managed edge computing platform.  However, since the unmanaged edge devices are not supervised by a particular entity, scheduling tasks to minimize latency in this scenario poses some unique challenges. These include substantial heterogeneity in computational capacity of the UEDs and task interference patterns among co-located tasks on one device, as well as runtime variations in the usable capacity of the edge devices. Also, the UEDs may only be available sporadically and have unpredictable churn. The existing scheduling schemes do not holistically consider all of these unique challenges and hence are not sufficient for task orchestration in an unmanaged setting. Also, many existing works~\cite{ad_hoc_1,Aral2019,GA,zenith} utilize the monitoring information provided by the edge devices such as CPU usage, frequency, memory usage, etc. to make decisions regarding which edge device a particular task should be offloaded to. In the case of unmanaged edge, this information may not be readily available due to privacy concerns by the owner of the device, the performance perturbation to collect such monitoring data, and the network cost of conveying that data. Even if it is available, with the amount of added dynamism and heterogeneity introduced by unmanaged edge, the information quickly becomes stale. 
% For example, a more powerful laptop may execute a task faster than a less powerful tablet, even if the current CPU usage is much higher for the laptop. A geographically closer computationally less powerful tablet may execute a task faster than a geographically farther more powerful laptop. Additionally, if we factor in the interference caused by co-located applications on a particular unmanaged edge device, decision making based on monitoring individual edge devices may be inaccurate or unscalable. 
\name overcomes these challenges and minimizes the service time and bandwidth overhead of tasks in the unmanaged edge scenario.

\begin{comment}

Another issue in making decisions based on the monitoring information in the case of unmanaged Edge is regarding the veracity of the information provided. The unmanaged Edge devices have an incentive to not provide the accurate monitoring information. For example, consider a volunteer-based incentive model, in which the incentive for a device to become part of the unmanaged Edge is that it can also use the unmanaged Edge to run its own applications. As shown in Figure 2 (a), in this scheme, if a particular unmanaged edge device falsely claims that all of its resources are exhausted, no tasks would be sent to that device while the device can still use unmanaged Edge to run its own tasks. In the case of a payment-based incentive model shown in Figure 2 (b), a device can falsely claim that it has a lot of resources available even when it is actually not. The device has an incentive to falsely provide the monitoring information so that more tasks are sent to the device thereby generating more payment. However, since the advertised resources are not actually available, the latency sensitive tasks would suffer. 

\begin{figure}[tb]
\centering
%\includegraphics[clip,trim=0cm 6cm 2.2cm 0cm,width=0.65\textwidth]{overview_diagram.pdf}
\includegraphics[width=\columnwidth]{false_info_incentive.png}
\vspace{-20pt}
\caption{“Unmanaged” Edge devices have an incentive to provide false monitoring information
}
\vspace{-13pt}
\label{fig:incentive_to_provide_false_info}
\end{figure} 

\end{comment}

\begin{figure}[t]
\centering
\includegraphics[width=\columnwidth]{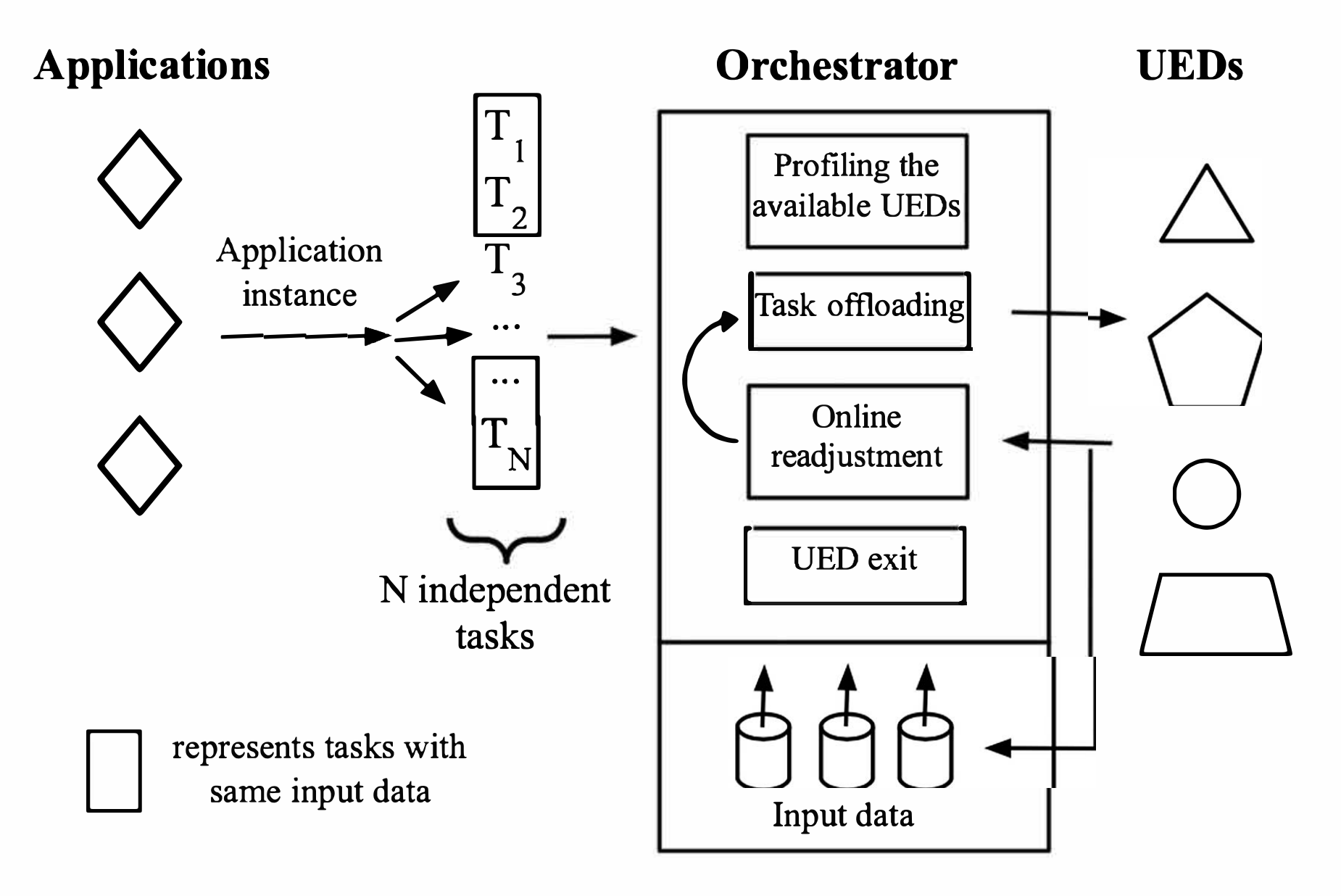}
\vspace{-20pt}
\caption{System Overview}
\vspace{-15pt}
\label{fig:system_overview}
\end{figure} 

\begin{comment}

Another issue in making decisions based on the monitoring information in the case of “unmanaged” Edge is the dynamic changes in the available resources (computational heterogeneity). The available resources of a $UED$ will vary depending upon the computational complexity of the personal applications being run by the owner of the $UED$. Hence, if monitoring information were to be used, it would be required to resend it repeatedly for it to be accurate. This would cause a lot of overhead in terms of bandwidth utilization.

For applications in which each user request consists of N independent tasks, an attractive method to minimize the total latency could be to send the N independent tasks to N different Edge servers. However, because of the dynamics involved in Edge Computing, a straightforward mapping of these tasks to the servers would be inefficient. A lot of factors need to be considered, such as: how far are the available Edge servers (network latency), what are their computational capabilities (computational latency), how loaded are the available Edge servers (interference latency), etc. Also, some of the tasks may pertain to the same input data. Sending all of these tasks to different Edge servers would be inefficient in terms of the bandwidth overhead, especially if the input data is huge.  

\end{comment}

Figure~\ref{fig:system_overview} presents an overview of the main components of \name. Each application instance from an end user consists of $N$ tasks, some of which are dependent in that they may require the same input data for execution. Application instances are sent to our orchestrator (\name) which schedules the tasks to the available heterogeneous UEDs. The orchestration scheme includes interference profiling of the available UEDs, selecting the optimal UEDs for the execution of tasks based on this profiling information and input parameters (such as the number of tasks running on the UEDs computed based on the number of tasks sent and responses received by the orchestrator), adjusting for the online heterogeneity based on the feedback and an efficient mechanism for UED exit\footnote{For simplicity of exposition, we describe the orchestrator as if it is centralized. In practice, standard fault-tolerance replication techniques can be used to make it distributed and fault-tolerant ~\cite{fault_tolerance_1,fault_tolerance_2}.}. The UEDs are selected to minimize the service time of the tasks and reduce the bandwidth overhead. Bandwidth overhead can occur if tasks that require the same input data are sent to different UEDs, especially if the input data is huge. \name does not require any monitoring information from the UEDs. 

In our evaluation, we compare \name to two intuitive baselines (random and round-robin assignment of tasks) and two state-of-the-art solutions, LAVEA~\cite{lavea} and Petrel~\cite{petrel}. Compared to the existing schemes, \name significantly reduces the average service time of application instances by at least $61\%$ (Figure~\ref{fig:running_avg_latency}). The reduction in average service time is more significant in the presence of online heterogeneities such as variation in the computational capability of UEDs (Figure~\ref{fig:variation_in_computation_capacity}) or sporadic availability of UEDs (Figure~\ref{fig:sporadic_availability}). At the same time, the bandwidth overhead for \name is at least $56\%$ lower than that of the other schemes (Figure~\ref{fig:bandwidth_overhead}).

% The main contributions of this paper are:

\noindent {\bf Contributions:} Our contributions in this paper can be summarized as follows. 

\begin{enumerate}[leftmargin=1.5em]
\addtolength\partopsep{-1em}
\addtolength\topsep{-1em}
% \addtolength\leftmargin{-2em}

% \item We propose the concept of “unmanaged” Edge to execute latency sensitive user application requests.

\item We present \name, an interference-based dynamic task orchestration scheme to execute user applications consisting of multiple tasks in a heterogeneous unmanaged Edge computing environment. \name optimizes for latency and bandwidth overhead in a configurable manner.

\item Our proposed orchestration scheme takes into consideration the heterogeneity in interference patterns across multiple $UEDs$, the sporadic availability of $UEDs$, and the runtime variations in their computational capacity due to co-located applications. It does not require any monitoring information from the UEDs.

\item We perform extensive simulations and real-world evaluations to demonstrate the effectiveness of \name over four baseline solutions.
  
\end{enumerate}

The rest of the paper is organized as follows: Section~\ref{sec:motivation_challenges} presents the motivation for unmanaged edge computing and the main challenges involved in task orchestration in this scenario. Section~\ref{sec:system_overview} provides a high-level overview of the system components. Section~\ref{sec:design} presents the design of the proposed orchestration scheme and Section~\ref{sec:evaluation} the evaluation results. Section~\ref{sec:discussion} elaborates on extensions to our work. Finally, Section~\ref{sec:related_work} discusses related work and Section~\ref{sec:conclusion} concludes the paper.

%% file: sec_motivation_challenges.tex
\section{Motivation and Challenges}
\label{sec:motivation_challenges}

\subsection{Motivating Example}
\label{subsec:motivating_example}

Consider a typical application from the domain of autonomous self-driving cars~\cite{autonomous_cars}. It has the tasks listed below and we use this application in our evaluation (one of three). 
% \begin{enumerate}[label=(\alph*)]
\begin{enumerate}
\item Driver state detection using face camera
\item Driver body position using driver cabin camera
\item Driving scene perception using a forward-facing camera
\item Vehicle state analysis using instrument cluster camera
\end{enumerate}

Task (c) can further consist of multiple tasks like pedestrian detection, obstacle detection, traffic signs analysis, etc. All these tasks would operate on the same input data, \ie the feed from the forward-facing camera.
In this paper, we focus on how to offload user requests pertaining to the latency-sensitive applications (such as the example above), in a heterogeneous unmanaged edge computing scenario. We aim at minimizing latency while providing a configuration parameter that determines how bandwidth conserving the allocation of tasks to UEDs is.

% \begin{figure*}[t]
% \centering
% \begin{multicols}{3}
%     \begin{subfigure}[t]{\columnwidth}
%     \includegraphics[width=0.66\columnwidth]{Figures/heterogeneity_distance.png}
%     \caption{}
%     \label{fig:Ng1} 
%     \end{subfigure}
%     \hfill
%     \begin{subfigure}[t]{\columnwidth}
%     \includegraphics[width=0.66\columnwidth]{Figures/heterogeneity_distance.png}
%     \caption{}
%     \label{fig:Ng2}
%     \end{subfigure}
    
%     \caption[]{}
    
%     \hfill

%     \includegraphics[width=0.66\columnwidth]{Figures/heterogeneity_distance.png}\par\caption{\label{fig:8}Comparison of Edge \& Cloud hybrid model with Cloud model}
%     \includegraphics[width=0.66\columnwidth]{Figures/heterogeneity_distance.png}\par\caption{\label{fig:9}Comparison of Mobile \& Edge hybrid model with Edge model}
% \end{multicols}
% \end{figure*}

\begin{figure}[t]
        \centering
        \begin{subfigure}[h]{0.49\columnwidth}
            \centering
            \includegraphics[width=\columnwidth]{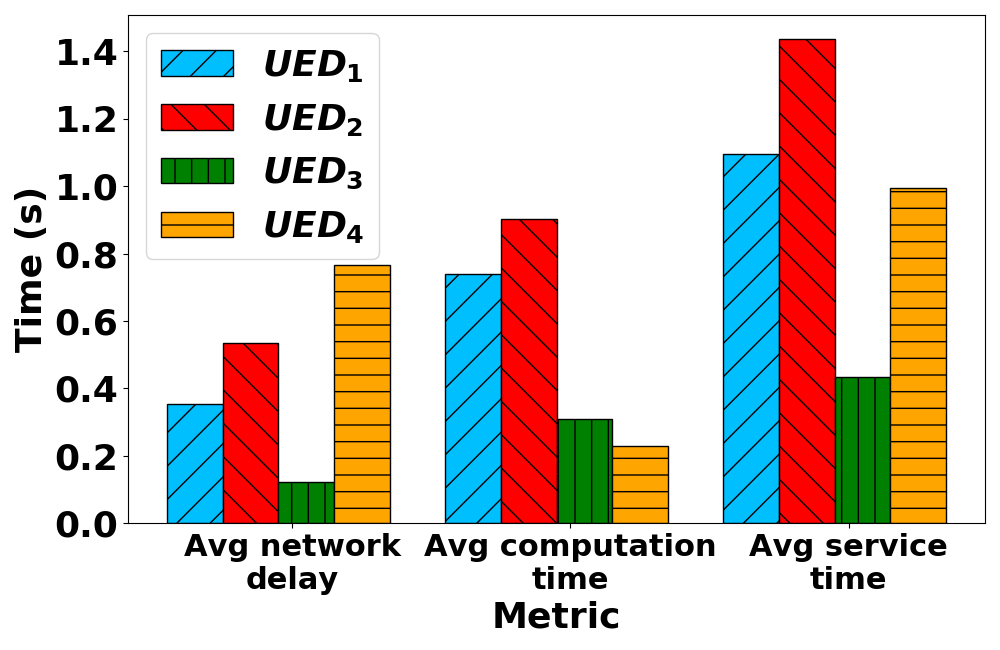}
            \caption[] {Computational and geographical heterogeneity}
            \label{fig:heterogeneity_computational}
        \end{subfigure}
        \hfill
        \begin{subfigure}[h]{0.49\columnwidth}  
            \centering 
            \includegraphics[width=\columnwidth]{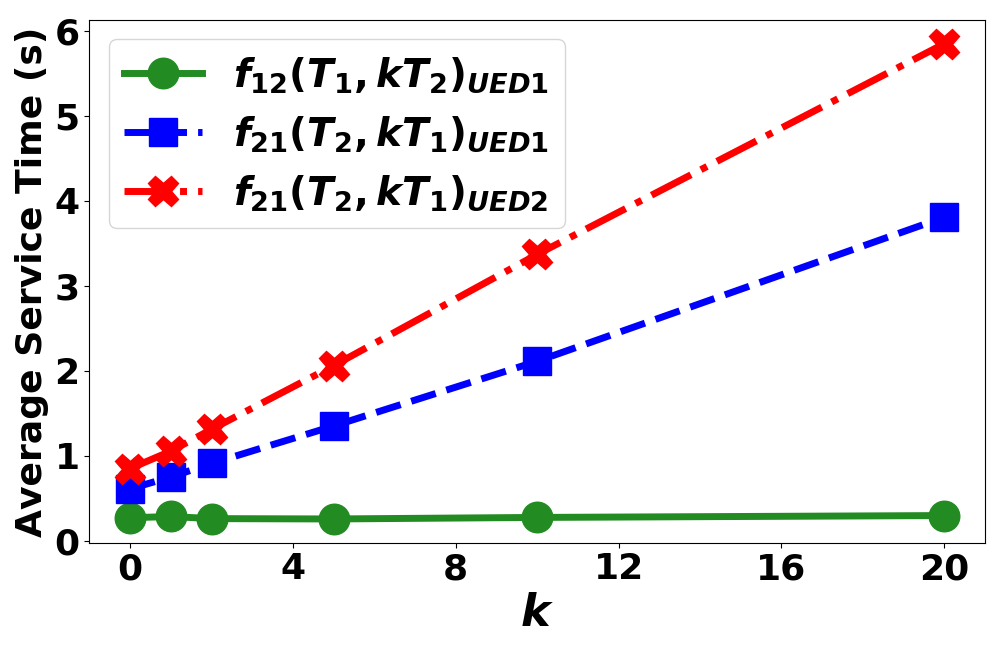}
            \caption[]{Heterogeneity in interference pattern}
            \label{fig:interference_heterogeneity}
        \end{subfigure}
        \vskip\baselineskip
        \vspace{-15 pt}
        \caption[]
        {Challenges in unmanaged edge orchestration} 
        \label{fig:challenges}
        \vspace{-15 pt}
\end{figure}

% \begin{figure}[t]
% \centering
% \includegraphics[width=\columnwidth, height = 5cm]{Figures/heterogeneity_distance.png}
% \vspace{-20pt}
% \caption{\label{fig:heterogeneity_computational}Computational and geographical heterogeneity}
% \centering
% \vspace{-5mm}
% \end{figure}

\subsection{Challenges and Responses}
\label{subsec:challenges}

The notion of unmanaged edge introduces a set of unique challenges unseen in traditional edge computing. Following are the main challenges involved in the orchestration of tasks in an unmanaged edge scenario and a brief statement about how we handle each challenge. 

    \noindent \textbf{Substantial heterogeneity in computational capacity and geographical distance of edge devices:} The edge devices, which are personal laptops, tablets, desktops, etc., in our case, consist of heterogeneous hardware and hence, the performance of a task varies significantly on different edge devices. Also, different edge devices are at different geographical distances from the orchestrator. Consequently, the network delay also varies. Figure~\ref{fig:heterogeneity_computational} shows the average service time (average network delay + average computation time) of executing an image classification task on four heterogeneous edge devices at varying distances from the orchestrator in a production setting. The four UEDs are Samsung Galaxy Tab S4-2018 ($UED_1$), Dell Inspiron 15R-2013 ($UED_2$), Macbook Pro-2018 ($UED_3$) and iMac-2017 ($UED_4$). Note the huge disparity between the average network delay (max-min ratio 6:1) due to geographical heterogeneity and the average computation time (max:min ratio 4:1) due to computational heterogeneity among the UEDs.

% \begin{figure}[t]%[t]
% \centering
% \includegraphics[width=\columnwidth, height=5cm]{Figures/task_interference.png}
% \vspace{-20 pt}
% \caption{\label{fig:interference_heterogeneity}Heterogeneity in interference pattern}
% \centering
% \vspace{-6mm}
% \end{figure}

% \begin{figure}
% \centering
% \begin{subfigure}[t]{\columnwidth}
% \centering
%   \includegraphics[width=\columnwidth, height = 4.5cm]{Figures/interference_T2_T1.png}
%   \caption[]{Interference of T1 on T2.}
%   \label{fig:interference_T2_T1}
% \end{subfigure}

% \begin{subfigure}[t]{\columnwidth}
%     \centering
%   \includegraphics[width=\columnwidth, height = 4.5 cm]{Figures/interference_T1_T2.png} 
%   \caption[]{Interference of T2 on T1.}
%   \label{fig:interference_T1_T2}
% \end{subfigure}

% \caption[]{Task interference on an edge device.}
% \label{fig:task_interference}

% \vspace{-5mm}
% \end{figure}  

% \begin{figure*}[t]
% \vspace{-4mm}
% \centering
% \includegraphics[width=2\columnwidth]{Figures/system_model.png}
% \vspace{-4mm}
% \caption{\label{fig:system_model} System model.}
% \centering
% \vspace{-2mm}
% \end{figure*}

    \noindent \textbf{Heterogeneity in task interference pattern:} Different tasks, when running on the same edge device, may interfere with each other affecting their service time. There is a heterogeneity in the interference experienced by different types of tasks on a UED. For instance, Figure~\ref{fig:interference_heterogeneity} considers task $T_1$, an image segmentation task, which is simpler compared to $T_2$, an image classification task. It shows the difference between the interference of tasks of type $T_1$ on $T_2$ \big($f_{21}(T_2,kT_1)_{UED1}$\big) and $T_2$ on $T_1$ \big($f_{12}(T_1,kT_2)_{UED1}$\big) on $UED_1$. The interference is quantified using $f_{ij}(T_i,kT_j)_{UEDp}$ which gives the execution time of a new task of type $T_i$ on $UED_p$, given that $k$ tasks of type $T_j$ are already running on the UED. It can be seen from the figure that there is a high interference of $T_1$ on $T_2$ but almost negligible interference of $T_2$ on $T_1$. Not only do different types of tasks interfere differently on the same device, but also there is variation in interference pattern across multiple devices. Figure~\ref{fig:interference_heterogeneity} shows the comparison between the interference of $T_1$ on $T_2$ on two different $UEDs$ \big($f_{21}(T_2,kT_1)_{UED1}$ and $f_{21}(T_2,kT_1)_{UED2}$\big). The interference of $T_1$ on $T_2$ is higher on $UED_2$ than that on $UED_1$. Thus, interference depends on the ordered pair of tasks and also the UED. \name performs a novel interference profiling of the UEDs to handle this heterogeneity in interference pattern (Section~\ref{subsec:adding_device}).

    \noindent \textbf{Online variations in the usable capacity of an edge device:} Depending upon the personal applications that the owner is running on a UED, the amount of resources available for edge services will vary. To prevent a slowdown of the UED, we need to reduce the usage of the device if the owner starts running a computationally demanding personal application. \name handles this using online readjustment based on a feedback mechanism (Section~\ref{subsec:online_readjustment}).

    \noindent \textbf{Lack of monitoring information from edge devices:} Most of the current edge orchestration schemes~\cite{ad_hoc_1,Aral2019,GA,zenith} utilize monitoring information, such as CPU usage, frequency, memory consumption, etc., from the edge devices to make offloading decisions. However, we do not use any such information because of the following reasons:
    
    \begin{enumerate}
    \item As the edge devices in our case are not managed by a single entity, the monitoring information may not be readily available. Also, the owners of the devices may be privacy sensitive about sharing such information with a third party. Note that they have signed up to contribute some compute resources to the unmanaged edge platform, but that can rarely be interpreted to mean that the device owners want the usage on their devices to be monitored.
    \item Monitoring a large number of edge devices with the level of frequency needed to be useful would result in a huge overhead. The devices would have to transmit monitoring information continuously as their usable capacity is susceptible to variations, due to co-located applications starting up and other factors that do not occur at a set frequency.
    \end{enumerate}
    In \name, the orchestrator learns from external observation and predicts the service time of tasks without using any monitoring information from the edge devices (Section~\ref{subsec:proposed_scheme}).

    \noindent \textbf{Sporadic availability of unmanaged edge devices:} 
    Unlike the traditional servers in a managed edge setting which are always available, the availability of an unmanaged edge device would depend upon the owner of that device. Hence, we cannot rely on the device being available for computation all the time. Depending upon the work pattern of the owner of a device, it may be available intermittently at different times of the day. Based on the history of the availability of UEDs, we predict their future availability and use it in our orchestration scheme (Section~\ref{subsec:availability_prediction}).

%% file: sec_system_overview.tex
\section{System Overview}
\label{sec:system_overview}

\begin{figure}[t]
\centering
\includegraphics[width=\columnwidth]{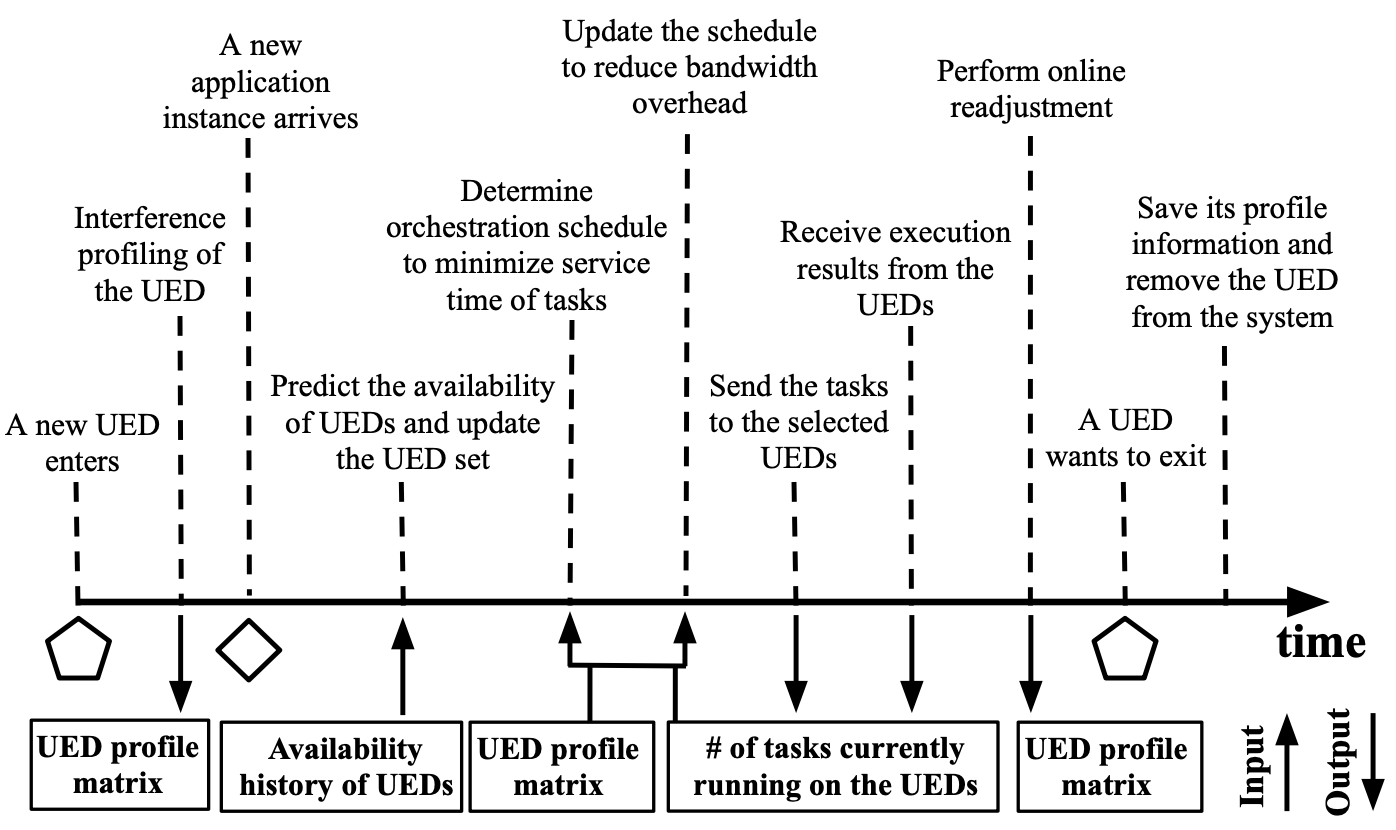}
\vspace{-20pt}
\caption{System Timeline}
\vspace{-10pt}
\label{fig:timeline}
\end{figure}

In this section, we present a high level overview of the main components of \name. Figure~\ref{fig:timeline} shows the timeline exhibiting the steps involved in adding a new UED to the system, orchestrating tasks to the available UEDs, performing online readjustment and gracefully removing a UED when it wishes to exit the system. As shown in Figure~\ref{fig:timeline}, when a new UED enters the system, our orchestrator profiles it using our novel interference-based profiling method (Section~\ref{subsec:adding_device}) and adds it to the UED profile database which stores the profiling information of all the added UEDs. This method of profiling handles the heterogeneity in the computational capabilities and interference patterns among the UEDs. When an application instance (consisting of $N$ different tasks) from an end user arrives at the orchestrator, the orchestrator first predicts which UEDs would be available throughout the execution of the application instance. It then updates the available $UED$ set to include only those UEDs which have a high probability of not leaving the system. This handles the sporadic availability of the UEDs, an inherent characteristic of unmanaged edge computing systems. An initial schedule for the $N$ tasks is then determined using the UED profile database and the data structure containing the number of tasks of different types already running on the available UEDs. This data structure is updated by the orchestrator whenever it sends a new task to a UED or receives an execution result from a UED. The initial schedule is a many-to-one mapping of the $N$ tasks to the available UEDs, aimed at minimizing the service time of the tasks. Next, \name updates the schedule to reduce the bandwidth overhead at the cost of a slight increase in the service time by trying to schedule the tasks that require the same input data on the same UED. \name includes a bandwidth overhead control parameter that manages this trade-off. The tasks are then sent to the selected UEDs. Upon receiving the execution results, the orchestrator sends them back to the end user. It then updates the UED profile database based on the error between the estimated and actual service time of the tasks on the selected UEDs. The error in the estimation of the service time can occur because of inaccurate profiling of a UED or online heterogeneity such as a variation in the available capacity of a UED. Updating the UED profile based on the feedback error handles such heterogeneities. In the event that a UED wishes to exit the system, its profiling information is saved by \name so that re-profiling is not required whenever the UED re-enters the system.

%% file: sec_design.tex
\section{Design}
\label{sec:design}

The system consists of our orchestrator running on a {\em managed} edge device that can offload tasks to multiple $UEDs$ connected to it, as shown in Figure~\ref{fig:system_overview}. The managed edge device is controlled by an infrastructure provider and can be a wireless access point, switch, low to mid range servers installed at the cellular base stations, etc. The end users send application instances to the managed edge device acting as the orchestrator. The orchestrator serves the instances in the order in which they arrive. Our goal is to minimize the total service time of all the tasks in the application instances while reducing the bandwidth overhead. 
% To handle the substantial heterogeneity in computational capacity, geographical distance and interference patters among the UEDs, we utilize a novel interference based profiling of the UEDs in our orchestration scheme. 
The symbols used in this paper and their definitions are summarized in Table~\ref{table:symbols_meanings}.

\begin{comment}
\begin{figure}[t!]
\centering
%\includegraphics[clip,trim=0cm 6cm 2.2cm 0cm,width=0.65\textwidth]{overview_diagram.pdf}
\includegraphics[width=\columnwidth]{arch_model.png}
\vspace{-20pt}
\caption{Architectural model}
%\vspace{-3pt}
\label{fig:incentive_to_provide_false_info}
\end{figure} 
\end{comment}

\subsection{Application Structure}
\label{subsec:application_strucutre}

Each application instance consists of $N$ tasks, some of which may require the same input data to execute. The structure of a typical application instance is shown in Figure~\ref{fig:system_overview}. It is more bandwidth efficient to send the tasks that require the same input data to the same UED. 
In our current implementation, we use a linear chain of tasks, though this can be extended to a DAG of tasks with no conceptual novelty (but some engineering effort), as discussed in Section~\ref{sec:discussion}.

\begin{figure}[t]
\centering
\includegraphics[width=\columnwidth, height=5cm]{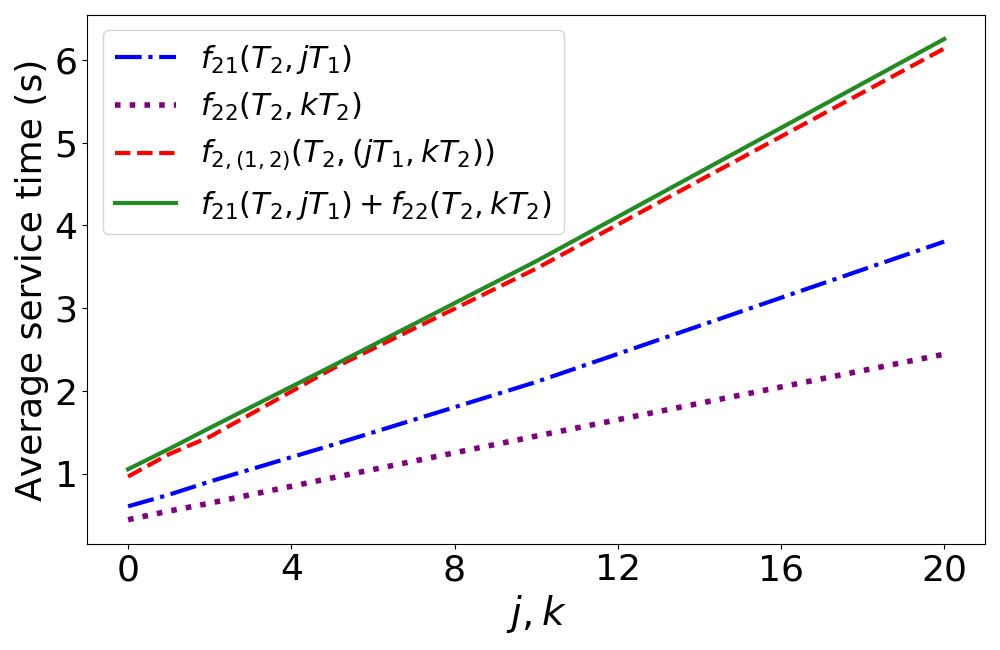}
\vspace{-8 mm}
\caption{\label{fig:incremental_validation}Experimental validation for computing the expected service time of a new incoming task using Eq.~\eqref{equation:incremental_validation}; $j$ and $k$ are the number of tasks of $T_1$ and $T_2$ already running on the UED respectively}
\centering
\end{figure}

\begin{table}
\footnotesize
\centering
\renewcommand{\arraystretch}{1.10}
\begin{tabular}{ |M{2.85cm}|M{4.9cm}|}
%  \hline
%  \multicolumn{2}{|c|}{\textbf{Symbols and their meanings}} \\
 \hline
 \textbf{Symbol} & \textbf{Definition}  \\[5 pt]
 \hline
 $T = \{T_1 , T_2, ... , T_N \}$ & $N$ different types of tasks for a given application instance \\
 \hline
 $UED = \{UED_1 , UED_2, ... , UED_Q \}$ & $Q$ is the total number of UEDs\\
 \hline
 $f_{ij}(T_i,kT_j)_p = m_{ij}*k + c_{ij}$ \hspace{5cm} $= <m_{ij}, c_{ij}>_p$  & Pairwise incremental service time plots on $UED_p$ characterized by slope $m_{ij}$ and y-intercept $c_{ij}$ \\
 \hline
 $A = [<m_{ij}, c_{ij}>_{p}]$ \hspace{15pt}  & Pairwise incremental service time matrix (each row corresponds to a different $UED$; Figure~\ref{fig:incremental_ST_matrix})\\
 \hline
 $Z = [z_{pi}] $ \hspace{60pt} & (Task count matrix) Number of tasks of type $T_i$ currently running on $UED_p$\\
 \hline
%  $S = [s_{pi}] $ \hspace{60pt} & Binary matrix that specifies whether the execution model for a task of type $T_i$ is present on $UED_p$\\
%  \hline
%  $L = [l_{pi}] $ \hspace{60pt} & Loading time of the execution model for a task of type $T_i$ on $UED_p$\\
%  \hline
 $ST_{exp}(T_i)_p$ \hspace{60pt} & Expected service time of a task of type $T_i$ on $UED_p$\\
 \hline
 $ST_{actual}(T_i)_p$ \hspace{60pt} & Actual service time of a task of type $T_i$ on $UED_p$\\
\hline
$R(t)_p$ \hspace{60pt} & Probability that $UED_p$ is available continuously between the current time and $t$ time units in the future\\
\hline
Hyper-parameters: \hspace{30pt}(i) $\delta$ (ii) $\beta$ (iii) $\gamma$ \hspace{60pt} & (i) $\delta$ controls the amount of readjustment performed online (ii) $\beta$ controls the amount of reduction in the bandwidth overhead (iii) $\gamma$ is minimum threshold for a UED availability for it to be used \\
 \hline
 
\end{tabular}

\begin{tablenotes}
\item $i,j \in [1:N]$ ; $p \in [1:N]$
\end{tablenotes}
\caption{Symbols and their definitions.}
\label{table:symbols_meanings}
\renewcommand{\arraystretch}{1}
\vspace{-22 pt}
\end{table}

\begin{figure*}
\begin{equation*}
A_{Q,N^2} = 
\begin{pmatrix}
<m_{11},c_{11}>_1 & \cdots & <m_{1N},c_{1N}>_1 & \cdots & <m_{ij},c_{ij}>_1 & \cdots & <m_{N1},c_{N1}>_1 & \cdots & <m_{NN},c_{NN}>_1 \\
<m_{11},c_{11}>_2 & \cdots & <m_{1N},c_{1N}>_2 & \cdots & <m_{ij},c_{ij}>_2 & \cdots & <m_{N1},c_{N1}>_2 & \cdots & <m_{NN},c_{NN}>_2 \\
\vdots  & \vdots  & \vdots & \vdots & \vdots  & \vdots & \vdots & \vdots & \vdots \\
<m_{11},c_{11}>_p & \cdots & <m_{1N},c_{1N}>_p & \cdots & <m_{ij},c_{ij}>_p & \cdots & <m_{N1},c_{N1}>_p & \cdots & <m_{NN},c_{NN}>_p \\
\vdots  & \vdots  & \vdots & \vdots & \vdots  & \vdots & \vdots & \vdots & \vdots \\
<m_{11},c_{11}>_Q & \cdots & <m_{1N},c_{1N}>_Q & \cdots & <m_{ij},c_{ij}>_Q & \cdots & <m_{N1},c_{N1}>_Q & \cdots & <m_{NN},c_{NN}>_Q \\
\end{pmatrix}
\end{equation*}
\caption{Pairwise incremental service time matrix $A$; $Q$ is the total number of UEDs and N is the total number of different types of tasks in each application instance}
\label{fig:incremental_ST_matrix}
\end{figure*}
\vspace{-10 pt}
\subsection{Pairwise Incremental Service Time Plots}
\label{subsec:incremental_curve}

We define pairwise incremental service time plots $f_{ij}(T_i,kT_j)_p$ to characterize the execution time of a new task of type $T_i$ on $UED_p$, given that $k$ tasks of type $T_j$ are already running on the UED. This captures the heterogeneity in the interference caused by the tasks. Examples of such plots can be seen in Figures \ref{fig:interference_heterogeneity} and ~\ref{fig:incremental_validation}. We observed that these plots are always straight lines but with varying slopes and y-intercepts due to the task interference and heterogeneity in interference patterns, as elaborated in Section~\ref{subsec:challenges}.
On a given UED, for a new task $T_i$, we can plot $N$ pairwise incremental service time plots, one for interference with every other type of task (including $T_i$). Hence, $N^2$ such plots exist for every $UED$ and we need to store only $N^2$ pairs of $m$ and $c$ values to characterize all the plots for that $UED$. We compute the expected service time of any new incoming task $T_i$ on $UED_p$, which has $\alpha_1, \alpha_2, \cdots, \alpha_N$ tasks of each type already running using the following equation:

\vspace{-13pt}

\begin{multline}
    f_{i,(1,2, \cdots, N)}\big(T_i,(\alpha_1T_1, \cdots ,\alpha_iT_i, \cdots, \alpha_NT_N)\big) = f_{i1}(T_i,\alpha_1T_1) + \\ \cdots + f_{ii}(T_i,\alpha_iT_i) + \cdots + f_{iN}(T_i,\alpha_NT_N).
    \label{equation:incremental_validation}
\end{multline}

This assumes that the interference patterns are independent and additive. We verify this experimentally as can be seen in Figure~\ref{fig:incremental_validation}. The figure shows that the curve obtained by adding $f_{21}(T_2,jT_1)$ and $f_{22}(T_2,kT_2)$ is very similar to $f_{2,(1,2)}(T_2,(jT_1,kT_2))$.

We define a pairwise incremental service time matrix $A$, each row of which contains the $N^2$ pairs of $m$ and $c$ values for a particular UED. See Figure~\ref{fig:incremental_ST_matrix} for the structure of matrix $A$. The element $<m_{ij}, c_{ij}>_p$ means that if we want to schedule a new task of type $T_i$ while $k$ instances of task $T_j$ are running on a $UED_p$, the service time of this task $T_i$ will be estimated as $m_{ij}*k + c_{ij}$. We also define a task count matrix $Z$, each row of which contains the number of tasks of all the different types currently running on a particular UED. Since the orchestrator sends the tasks and receives the execution results from the UEDs, it keeps updating the matrix $Z$, whenever needed. Note that, in practice, the application instances arriving at the orchestrator will not be of the same application type. The application instances can be of different types, each consisting of a different set of tasks. At the orchestrator, there will be a separate matrix $A$ for each application type. However, for ease of exposition, we will present our algorithms as if all application instances that arrive belong to a single type of application consisting of $N$ tasks.

\begin{comment}

\subsection{State of Task Execution Models}
\label{subsec:state_tasks}

The execution of a particular type of task on a UED would require that the respective task execution model is already loaded on that device. For instance, an image classification task would require a deep neural network model. If the model is not present on the device, the total service time would include an additional model loading time apart from the execution time. Once the model for a particular type of task is loaded on the device, any future tasks of that type can be executed without the loading delay, as long as the model is still available on the device. The model may be removed from a device after it is loaded if the available memory space runs out and a different model for another task needs to be loaded.  

We define a binary matrix $S$, which specifies whether the execution model for different types of tasks is available on the UEDs. We also define a matrix $L$ containing the loading time for the execution models of different types of tasks on the UEDs. 

\end{comment}

\subsection{Interference Profiling: Adding a New Unmanaged Edge Device}
\label{subsec:adding_device}

Adding a new UED to the system requires obtaining all the $N^2$ pairs of $m$ and $c$ values for the UED and adding them as a new row to matrix $A$ (Figure~\ref{fig:incremental_ST_matrix}). One way to obtain the $N^2$ pairs is to recreate all the required pairwise interference patterns by actually running tasks on the UED. Since each pairwise interference pattern is a straight line, the $m$ and $c$ values for that pattern can be obtained by extracting any two points on the plot. However, this method of profiling a new UED is not desirable for large $N$ since it would require a lot of time and resources to obtain all $N^2$ pairs. For some UEDs, the amount of time needed to profile may be in the order of several minutes. Also, since the availability of UEDs in the unmanaged setting is sporadic, spending a lot of time in profiling a UED would be inefficient if the UED is not available for long.

To quickly profile a new UED, we use a technique similar to ~\cite{paragon}, which relies on Singular Value Decomposition (SVD) and PQ reconstruction. This technique is based on the algorithm Netflix uses to provide movie recommendations to new users who have only rated a handful of movies. The idea is to find similarities between the new user and the existing users who have rated a lot of movies. We profile the first few UEDs by actually obtaining all the $N^2$ pairs. Thereafter, for every new UED, we get as many pairs as possible within a fixed time bound (1 minute in our experiments and configurable) and estimate the missing pairs using SVD and PQ-reconstruction. The time complexity of SVD and PQ-reconstruction is linear in $N$ and, in practice, only takes a few milliseconds even for a large $N$ ($\sim$ 30). Hence, this scheme is much quicker than obtaining all the $N^2$ pairs. The inaccuracies in the estimation are handled by online readjustment (Section~\ref{subsec:online_readjustment}).

\subsection{UED Availability Prediction}
\label{subsec:availability_prediction}

One of the challenges in unmanaged edge computing is the sporadic availability of the UEDs (Section~\ref{subsec:challenges}). UEDs may enter or exit the system without prior notice. If a task is scheduled on a UED which is unavailable, or which exits the system before task completion, it would be required to reschedule the task thereby increasing the task completion time. \name predicts the availability of the UEDs and schedules tasks on a UED only if there is a high probability of it being available throughout the task completion. We utilize a semi-Markov Process (SMP) model, similar to ~\cite{availability_prediction}, to predict the reliability $R$ of a UED. This is the probability of the UED being available throughout a future time window. In an SMP model,  the  next transition not only depends on the current state (as would happen for a pure Markov model) but also on how  long  the  system  has  stayed  at  this  state. We observed that the availability pattern of a UED is comparable in the most recent days. Hence, using the availability history of a UED on previous days, we calculate the parameters of the SMP to evaluate $R(t)$, the probability that the UED is available continuously between the current time and $t$ time units in the future. Tasks are scheduled on a UED only if the probability of it being available throughout the time that it takes to complete the most demanding task in the application is greater than a threshold $\gamma$.

\subsection{Orchestration Scheme}
\label{subsec:proposed_scheme}

The orchestration algorithm, the largest part of \name, is shown in Algorithm~\ref{algo:main_orchestrator}. The algorithm consists of four segments: UED availability prediction, minimum service time scheduling, reduction in the bandwidth overhead, and online readjustment. When a new application instance arrives, we first predict the probability of each UED being available throughout the execution of the application instance. The UEDs for which this probability is lower than a threshold $\gamma$ are dropped out of the scheduling for the current application instance. The orchestrator maintains a count (in matrix $Z$) of the number of tasks of different types currently running on the available UEDs. The orchestrator uses this count and the pairwise incremental service time matrix $A$ to predict the service time of the tasks on every available UED and create an initial mapping between the tasks and the UEDs. This mapping assigns each task to a UED on which the expected service time for the task is minimum under the current state of other tasks running on each UED. 
Predicting the service time of a task involves extracting the corresponding entries from the A matrix and using Eq.~\ref{equation:incremental_validation}. 
% Algorithm~\ref{algo:get_expected_ST} shows how to predict the service time of Task $T_i$ on $UED_p$. It involves extracting the required $<m, c>$ values for $T_i$ and $UED_p$ from $A$ and the number of different types of tasks currently running on the $UED_p$ from $Z$. Equation~\ref{equation:incremental_validation} can then be used to get the expected service time.

Next, the orchestrator tries to reduce the bandwidth overhead by making modifications to the initial schedule. For every group of tasks that require the same input data but are scheduled on different UEDs, the orchestrator tries to schedule them on the same UED to reduce the bandwidth overhead. A change in the assigned UED for a task is made only if the relative increase in its service time due to the change is less than a threshold $\beta$, which is the bandwidth overhead control parameter. It decides the trade-off between the bandwidth overhead and the average service time. If $\beta$ is higher, \name becomes more bandwidth conserving at the expense of higher service time. Finally, the tasks are sent and executed by the assigned UEDs. Upon receiving the execution result, the orchestrator computes the actual service time for each task. If the difference between estimated and actual service times for a task is greater than an error threshold ($\delta$), then the orchestrator updates $A$ as described  (Section~\ref{subsec:online_readjustment}). For $Q$ total number of UEDs and $N$ tasks in each application instance, the time complexity of our orchestration scheme is $\mathcal{O}(NQ)$. Hence, our scheme can easily scale up without significant overheads.

% The orchestration scheme also considers the cost of loading a new model on the UEDs. If the model required to execute task $T_i$ is unavailable on $UED_p$, the expected service time includes the loading time in addition to the expected computation time. 

\begin{algorithm}[ht!]
\footnotesize
\SetAlgoLined
\textbf{Input:} A new application instance $T$\\
\textbf{Initialization:} $UED$, $A$ and $Z$ \\
 
%  Let $T_c$ be the most computationally intensive task \\
%  \For{$UED_p \in UED$}{
%   $t_p = GetExpectedServiceTime(c, p)$ \;
%   }
%   $t_{max} = \max\limits_{p}\big(t_p\big)$ \;
Let $t_{max}$ be the maximum time to execute the most computationally intensive task on the devices in $UED$

 \DontPrintSemicolon
 \hspace{0.1cm}\tcp{UED availability prediction}
 \PrintSemicolon
 \For{$UED_p \in UED$}{
 Compute $R_p(t_{max})$ using semi-Markov Process (SMP)\\
 \If{$R_p(t_{max}) \leq \gamma$}{
        Remove $UED_p$ from $UED$\\
    }
 }
 
%  Predict the probability of the devices in $UED$ being available at the end of $t_{max}$ units of time \\
 
%  Remove from $UED$, the devices with availability prediction probability < $\gamma$

\DontPrintSemicolon
\hspace{0.1cm}\tcp{Minimum service time scheduling}
\PrintSemicolon
 
 \For{$T_i \in T$}{
  \For{$UED_p \in UED$}{
  $ST_{exp}(T_i)_p =  GetExpectedServiceTime(i, p)$ \;
  }
  
  \vspace{1pt}
  
  $ST_{exp}^{min}[i] = \min\limits_{p}\big(ST_{exp}(T_i)_p\big)$ \;
  
  $UED_{sel}[i] = \argmin\limits_{p}\big(ST_{exp}(T_i)_p\big)$ \;
  }
  
\DontPrintSemicolon
\hspace{0.1cm}\tcp{Reduction in bandwidth overhead}
\PrintSemicolon
 
  Let $K = [k_1, k_2, ... k_R]$ be a group of tasks which require the same input data
  
  \For{every $K$}{
  $ued_{1} = UED_{sel}[k_1]$ \;
  \For{j = 2, ..., R}{
    $ued_j = UED_{sel}[k_j]$\;
    \If{$ued_j \neq ued_{1}$}{
    $ST_{min} = ST_{exp}^{min}[k_j]$\;
    $ST_{1} = GetExpectedServiceTime(k_j, ued_1) $\;    
    \If{$\frac{ST_{1} - ST_{min}}{ST_{min}} \leq \beta$}{
        $UED_{sel}[k_j] = ued_1$\;
    }
   }
 }
}
 
\DontPrintSemicolon
\hspace{0.1cm}\tcp{Online readjustment}
\PrintSemicolon
  
  \For{$T_i \in T$}{
  $p = UED_{sel}[i]$ \;
  Schedule task $T_i$ on $UED_p$ and compute the actual service time $ST_{actual}(T_i)_p$\
%   $ST_{actual}(T_i)_p = GetActualServiceTime(i, p)$ \;
  
  $ST_{exp}(T_i)_p = ST_{exp}^{min}[i]$\;
  
  \If{$\frac{\big|ST_{exp}(T_i)_p - ST_{actual}(T_i)_p \big|}{ST_{actual}(T_i)_p} > \delta$}{
   \vspace{2pt}
   $PerformGradientDescent\big(i,p,ST_{exp}(T_i)_p,$ \\\ \hspace{4 cm} $ST_{actual}(T_i)_p\big)$ \;
   }
 }
 \caption{$Main\_Orchestrator$}
 \label{algo:main_orchestrator}

\end{algorithm}

\vspace{-7 pt}

\begin{comment}
  
\begin{algorithm}[t]
\SetAlgoLined
\DontPrintSemicolon
\textbf{Input:} $i, p$\\
\textbf{Output:} Actual Service Time for a Task of Type $T_{i}$ on $UED_p$\\
 
 \PrintSemicolon
 $start\_time = GetCurrentTime()$\;
 Send $T_i$ to $UED_p$\;
 Receive the task execution result from $UED_p$\;
 $end\_time = GetCurrentTime()$\;
 $ST_{actual}(T_i)_p = end\_time - start\_time$\;
 return $ST_{actual}(T_i)_p$;
 
 \caption{$GetActualServiceTime(i, p)$}
 \label{get_actual_ST}
\end{algorithm}

\end{comment}

\begin{comment}
  Also, the orchestrator based on a history of availability of different Edge Devices, predicts the future availability of the connected Edge Devices and does not schedule a task on an Edge Device if it cannot be completed before the Device disconnects.
\end{comment}

\subsection{Online Readjustment}
\label{subsec:online_readjustment}

Online readjustment of the $p^{th}$ row of matrix $A$ is needed when there is a large difference (greater than $\delta$) between the expected and the actual service time of a task $T_i$ on $UED_p$. This difference arises if there is an inaccuracy in the $N$ incremental service time pairs $<m , c>$ corresponding to $T_i$ in the $p^{th}$ row of $A$. Following are the main reasons for the inaccuracy:

    \noindent \textbf{Imperfect information:} As described in Section~\ref{subsec:adding_device}, most of the $<m , c>$ pairs in the row added for a UED are computed using SVD and PQ reconstruction and may not be completely accurate. 
    
    \noindent \textbf{Online variation:} Even if all the $<m , c>$ pairs are correctly profiled initially, the true values may change over time if the owner of the UED starts using a larger portion of the device's compute capability for his/her personal applications. This will result in a change in the pairwise incremental service time plots, thereby changing the $<m , c>$ values.

Therefore, we need to make online adjustments to the matrix A. For this, we use gradient descent as described in Algorithm~\ref{algo:perform_GD}. For a task $T_i$ scheduled on $UED_p$, if the difference between the expected and the actual service time exceeds $\delta$, gradient descent is performed to minimize the error between the expected and actual service time and obtain the new values of $<m , c>$ for task $T_i$ on $UED_p$.

\subsection{Unmanaged Edge Device Exit}
\label{subsec:edge_device_exit}

A UED may leave the system if there is a sudden unexpected crash or if the owner of the UED exits the system. Not much can be done in the case of an unexpected crash. However, in the other case, we perform an additional step for a graceful exit which can save us from re-profiling the UED if it re-joins the system in the future. When the owner of the $UED_p$ wants to exit the system, the information corresponding to the UED stored in the $p^{th}$ row of the $A$ matrix is saved by the system. The row can then be removed from $A$ in the orchestrator. Later, if the UED rejoins the system, its profiling information can be loaded to the orchestrator during the entry phase which significantly reduces the time needed to profile the UED on the system. Concurrently, any needed model or data can be loaded on to the rejoining UED through an efficient wireless reprogramming protocol~\cite{panta2011efficient}. 
 
% SB (7/10/20): Chopped for space 
\begin{comment}
\begin{algorithm}[t]
\footnotesize
\SetAlgoLined
\DontPrintSemicolon
\textbf{Input:} $i, p$\\
\textbf{Output:} Expected Service Time for a Task of Type $T_{i}$ on $UED_p$\\
 
 $\xv = TaskCountUED_p = Z[p,:] = [Z_{pj}] ; j \in [1:N]$\
 \hspace*{4cm}\tcp*{$\mbox{row}_p(Z)$}
 
 \PrintSemicolon
 
 $\mv = [<m_{ij}>_p]$ ; $j \in [1:N]$\ \\
 $\cv = [<c_{ij}>_p]$ ; $j \in [1:N]$\ 
 \DontPrintSemicolon
 \\
\tcp*{$\mv$ and $\cv$ extracted from $\mbox{row}_p(A)$}
 
 $ST_{exp}(T_i)_p = \mv\xv^{T} + \onev\cv^T$\;
 
 return $ST_{exp}(T_i)_p$;
 
 \caption{$GetExpectedServiceTime(i, p)$}
 \label{algo:get_expected_ST}
\end{algorithm}
\end{comment}

\begin{algorithm}[t]
\footnotesize
\SetAlgoLined
\DontPrintSemicolon
\textbf{Input:} $i, p, ST_{exp}, ST_{actual}$\\
 \PrintSemicolon
 $M = [<m_{ij}>_p]$ \;
 $C = [<m_{ij}>_p]$ ; $j \in 1, 2, ..., N$\ 
 \DontPrintSemicolon
 \\
\tcp*{$M$ and $C$ extracted from $p^{th}$ row of $A$}
 $X = TaskCountUED_p = Z[p,:] = [Z_{pj}] ; j \in 1, 2, ..., N$\
 \hspace*{4cm}\tcp*{$p^{th}$ row of $Z$}
 \PrintSemicolon
 
 $M^{new}, C^{new} = GradientDescent(M, C, X, ST_{actual},$   $ST_{exp})$\;
 
 Update $A$ with $M^{new}$ and $C^{new}$;

\caption{$PerformGradientDescent(i, p, ST_{exp}, ST_{actual})$}
\label{algo:perform_GD}
\end{algorithm}

%% file: sec_evaluation.tex
\section{Evaluation}
\label{sec:evaluation}

In this section, we first present the major findings from a survey conducted to understand the feasibility of unmanaged edge computing. Then, we provide details about the real-world experiments and how we used them to drive our simulations. We compared the service time obtained by \name with two baseline schemes and two state-of-the-arts for a latency sensitive application. The aim was to evaluate whether our scheme reduces the average service time of the application without a considerable increase in the bandwidth and the orchestration overhead. 
We then performed a set of micro experiments to evaluate the effect of various control parameters. 
% To thoroughly evaluate all the aspects of our design, we performed simulations with online readjustment and analyzed the impact of online heterogeneties such as sporadic availability of the $UEDs$ and changes in their computation capacity. We also compared the bandwidth and orchestration overheads as well as the fairness of the orchestration schemes. 
We then show that our solution works for two other applications, of light and medium load compared to the autonomous driving one introduced earlier. 
% Finally, we performed micro evaluations to understand the impact of varying our design parameters on the performance.

\begin{table}[b]
\footnotesize

\centering
\renewcommand{\arraystretch}{1.20}

\begin{tabular}{ |M{2.20cm}|M{2.25cm}|M{2.78cm}|}
 \hline
 \multicolumn{3}{|c|}{\textbf{Application Type}} \\

 \hline
 \textbf{Light} & \textbf{Medium} & \textbf{Heavy} \\[5 pt]
 \hline
 
color detection \hspace{30pt} $0.06 s$  \hspace{50pt} image segmentation \hspace{30pt} $0.12 s$  \hspace{50pt} edge detection \hspace{30pt} $0.17 s$ & kernel filtering \hspace{30pt} $0.22 s$ \hspace{50pt} contour detection \hspace{30pt} $0.25 s$ \hspace{50pt} feature transformation \hspace{30pt} $0.35 s$ & driver state detection $0.39 s$ \hspace{50pt} driver body position $0.45 s$ \hspace{50pt} vehicle state analysis $0.43 s$ \hspace{50pt} pedestrian detection $0.57 s$ \hspace{50pt} obstacle detection $0.60 s$ \hspace{50pt} traffic sign analysis $0.41 s$ \\
 
 \hline
 
\end{tabular}

\caption{Average service time of tasks for different application types}
\label{table:different_applications}
\renewcommand{\arraystretch}{1}
\vspace{-22 pt}
\end{table}

\subsection{Feasibility of Unmanaged Edge: A Survey}
\label{subsec:survey}

\begin{figure*}[ht]
        \centering
        \begin{subfigure}[h]{0.66\columnwidth}  
            \centering 
            \includegraphics[width=\columnwidth]{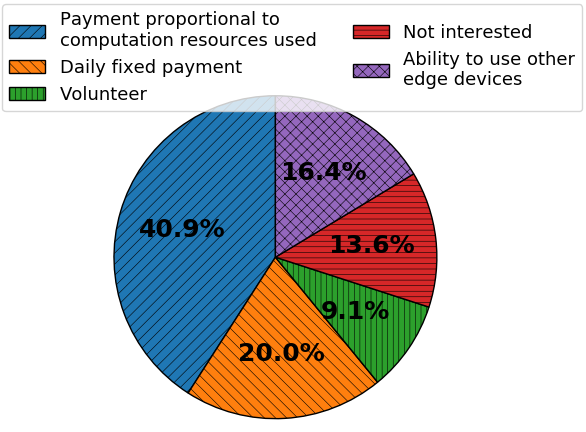}
            \caption[]{Preferred incentive model}
            \label{fig:incentive_model}
        \end{subfigure}
        \hfill
        \begin{subfigure}[h]{0.66\columnwidth}  
            \centering 
            \includegraphics[width=\columnwidth]{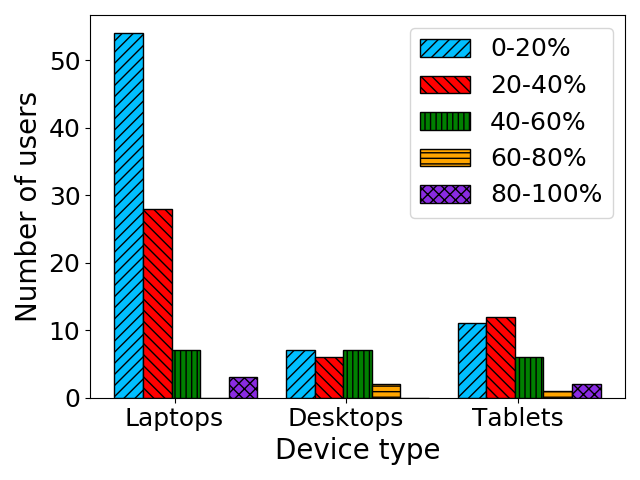}
            \caption[]{Percentage of CPU resources willing to share}
            \label{fig:resource_shared}
        \end{subfigure}
        \hfill
        \begin{subfigure}[h]{0.66\columnwidth}
            \centering
            \includegraphics[width=\columnwidth]{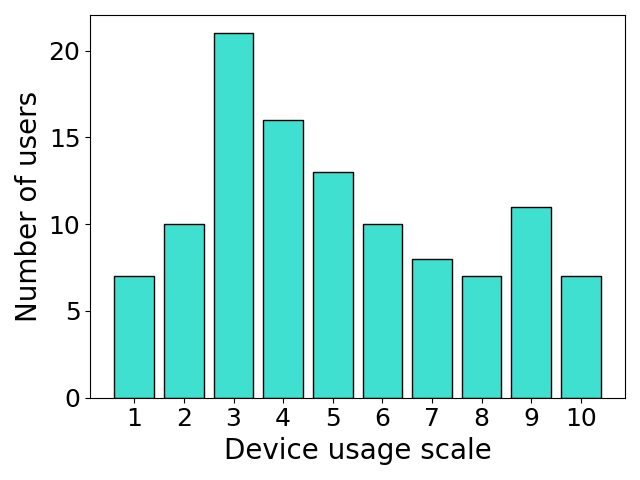}
            \caption[] {Device usage (1: Low usage, 10: Very high usage)}
            \label{fig:device_usage}
        \end{subfigure}
        \vspace{-5 pt}
        \caption[]
        {User survey results ($N$ = 110)} 
        \label{fig:survey_results}
        \vspace{-5 pt}
    \end{figure*}

We surveyed 110 participants --- from USA and India engaged in diverse fields such as educators, software professionals, students,  engineering professionals, etc. --- to understand the feasibility of unmanaged edge computing. $86.4\%$ of the participants indicated their willingness to provide their computing devices (\eg laptops, desktops, tablets, etc.) as UEDs under one of four proposed incentive models. Only $13.6\%$ of the participants were not interested primarily because of privacy and security concerns. The major takeaways from the survey are the following, as shown in Figure~\ref{fig:survey_results}:

% The primary purpose of the survey was to determine what percentage of participants would be willing to provide their computing devices (e.g., laptops, desktops, tablets, etc.) as edge devices. We discovered the preferred incentive model for unmanaged edge computing, average resource utilization of a typical user, and the percentage of resources he/she would be willing to share. We also identified how much slowdown the participants would be ready to tolerate as a result of their devices being used for edge computation.

\noindent \textbf{Preferred incentive model:} As expected, the majority ($40.9\%$) of the participants were willing to contribute their devices for edge computing if they received a payment proportional to the computational resources of their devices used, as shown in Figure~\ref{fig:incentive_model}. Daily fixed payment ($20\%$) and the ability to use other Edge devices for their applications ($16.4\%$) were second and third most popular choices respectively. 

% $9.1\%$ of the participants indicated that they would volunteer with no payment expected.

\begin{comment}

\noindent \textbf{Number of devices owned:} Laptop was the most popular device with almost everyone owning at least one laptop. A considerable number ($30\%$) of participants even owned two laptops. $35\%$ of the participants had at least one tablet, and $22\%$ had at least one desktop, as shown in Figure~\ref{fig:num_devices}. 

\end{comment}

\noindent \textbf{Percentage of CPU resources willing to share:} Most of the participants were willing to share between $0 - 40\%$ of the CPU resources of their devices, as shown in Figure~\ref{fig:resource_shared}. It is interesting to note that for tablets, more people were willing to share $20-40\%$ resources as compared to laptop owners who mostly showed a willingness to share $0-20\%$ resources. This result indicates that people do not use the computational resources of tablets as extensively as laptops and are willing to share more resources of their tablets. 
    
\noindent \textbf{Device usage and tolerable slowdown:} As shown in Figure~\ref{fig:device_usage}, we obtained a double Gaussian device usage pattern with peaks at $90\%$ and $30\%$ of usage indicating that most people either use their devices very heavily (video editing, running sophisticated software, etc.) or use them only for minor purposes such as browsing, reading, etc. The 
% peak at $30\%$ was stronger and the 
average usage across all users was $50.9\%$, thereby supporting our claim that a lot of devices are not utilized to their capacity. The majority of the people indicated that they could tolerate around $30\%$ slowdown of their devices.

\begin{comment}

\subsection{Typical Scenarios for Unmanaged Edge}

At present, the clients in all the cases below, transmit the data to the cloud for processing. Instead of adding new Edge infrastructure, can we use the power of already existing laptops/desktops/servers in the vicinity as potential Edge devices?

Elaborate on the following scenarios:

\begin{itemize}

    \item Client: Video camera at the traffic signal.
Unmanaged Edge: Laptops/Desktops in the vicinity
Application: Video surveillance

\item Client: cars
Unmanaged Edge: Laptops/Desktops in the vicinity
Application: Self-driving autonomous cars

\item Application: Augmented Reality (Virtual fitting room / virtual tour of a store)

\end{itemize}

\end{comment}

\subsection{Real-World Experiment}
\label{subsec:real_world}
The purpose of this experiment was to ensure that our simulations (described later) are based on real-world application and device data. We performed experiments with 3 different application types: a light-weight, a medium, and a heavy application. The tasks in each application type and their average service time on a typical UED (the Macbook Pro one in our testbed) are given in Table~\ref{table:different_applications}. For instance, the heavy application is the autonomous self-driving car application as described in Section~\ref{subsec:motivating_example}. This application consists of 6 tasks, 3 of which require the same input data. We obtained the incremental service time curves on 15 heterogeneous $UEDs$ (laptops, desktops and tablets) by running actual application instances on the $UEDs$. For example, the incremental service time curves shown in Figures~\ref{fig:interference_heterogeneity} and ~\ref{fig:incremental_validation} were obtained by running real tasks on actual $UEDs$. We used this data to drive the simulations to ensure that our simulation results are representative of reality.

\subsection{Simulation Setting}

Our simulator, built in Python, considers Poisson arrival of application instances with rate $\lambda$. Unless otherwise mentioned, we performed simulations with $500$ arrivals of the heavy application instances and set the default values of our hyperparameters for the experiments to $\lambda = 3$ $arrivals/s$, $\delta = 0.10$, $\beta = 0.15$, and $\gamma = 0.85$. For every arrival, the orchestrator needs to schedule the 6 tasks among the 15 available $UEDs$. We also provide a comparison for different application types in Section~\ref{subsec:different_application_types}. We compared our scheme with the following orchestration schemes:

\noindent \textbf{LAVEA:} Proposed in ~\cite{lavea}, LAVEA is a system that offloads computation to edge devices, to minimize the service time for low-latency video analytics tasks. They propose multiple task placement schemes for collaborative edge computation. We compare with their Shortest Queue Length First (SQLF) scheme, which performed the best among their task placement schemes in our evaluations. It tries to balance the total number of tasks running on each edge device.

 \noindent \textbf{Petrel:} Proposed in ~\cite{petrel}, Petrel is a distributed task scheduling framework for edge, which employs the strategy of "the power of two choices"~\cite{power_of_two}. In this scheme, two of the available edge devices are randomly selected and the task is sent to the one with lower expected service time. They compute the expected service time using the processor speed of the available $UEDs$.

 \noindent \textbf{Round Robin:} In this scheme, the tasks of an application instance are sent to the available $UEDs$ one after the other, i.e., the first task is sent to $UED_1$, the second to $UED_2$, and so on.
 
 \noindent \textbf{Random:} This is the most basic scheme in which each task of an application instance is sent to a randomly chosen $UED$.

We evaluated two versions of our scheme: 

\noindent \textbf{\nameP:} This stands for I-BOT-Perfect-Information. In this scheme, all the $UEDs$ are correctly profiled by obtaining the $N^2$ (36) pairs of $m$ and $c$ values for every $UED$. As mentioned in Section~\ref{subsec:adding_device}, this may not be desirable in the real-world because of high initialization time and sporadic availability of $UEDs$. This can be considered as the ideal case with lowest service time against which we compare the other schemes.

\noindent \textbf{\nameI:} This stands for I-BOT-Imperfect-Information. In this scheme, we use SVD and PQ reconstruction to create the incremental service time matrix $A$ (Section~\ref{subsec:adding_device}). This profiling is faster and realistic, but the incremental service time matrix so obtained may have inaccuracies, which are handled using online readjustment.

In our evaluation, we used the following \textbf{performance metrics:}

\noindent \textbf{Service Time:} For an application instance scheduled by the orchestrator, we define service time as the average completion time of the different tasks in the application instance. We define average service time (service time averaged over all the instances) and running average service time (service time averaged over a moving window of $50$ instances). For experiments in which there is a high fluctuation in the service time of application instances, plotting individual service time hinders visualization. We use running average service time for such experiments.

\noindent \textbf{Orchestration Overhead:} We define orchestration overhead for an application instance as the total amount of time spent by the orchestration scheme to decide where to schedule the instance. The average over all the application instances is defined as the average orchestration overhead. 

\noindent \textbf{Bandwidth Overhead:} We define bandwidth overhead for an application instance as the percentage of tasks that require the same input data but are sent to different $UEDs$. The average over all the application instances is defined as the average bandwidth overhead.

\subsection{Evaluation of the Orchestration Schemes}

\begin{figure}[t]
\centering
\includegraphics[width=\columnwidth]{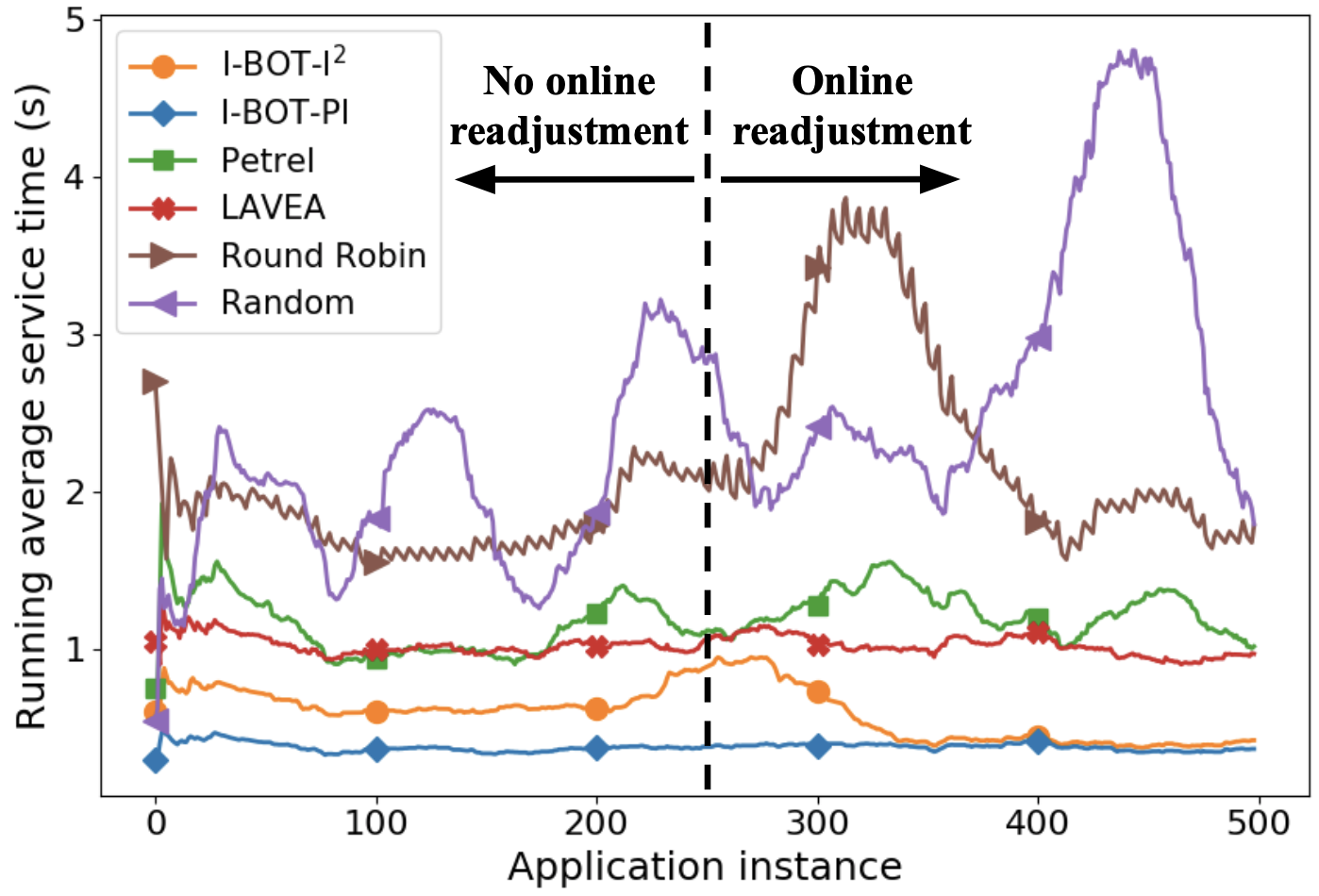}
\vspace{-5 mm}
\caption{\label{fig:running_avg_latency} Comparison of running average service time for different orchestration schemes}
\centering
\vspace{-15 pt}
\end{figure}

In this experiment, we compare the running average service time obtained by different orchestration schemes for 500 application instances arriving at rate $\lambda$ = 3 instances/sec. It can be observed from Figure~\ref{fig:running_avg_latency} that the service time for our schemes is significantly lower than that for the others. Note that, as mentioned earlier, \nameI would require online readjustment. To show the impact of online readjustment, we have used online readjustment only in the right half of Figure~\ref{fig:running_avg_latency} (from application instance $250$ onwards). For the left half, i.e, without online readjustment, the average service time for \nameP and \nameI are $0.39 s$ and $0.72 s$ respectively. This is $61.39\%$ and $28.71\%$ lower than the next best scheme LAVEA for which the average service time is $1.01 s$. Our performance is better because our schemes take into consideration the different interference patterns of tasks across the $UEDs$, which is not considered by the others. For instance, consider two $UEDs$ ($UED_1$ and $UED_2$) such that there is a high interference between tasks of type $T_1$ and $T_2$ on $UED_1$ and low interference on $UED_2$. In this scenario, our schemes will refrain from concurrently scheduling tasks of type $T_1$ and $T_2$ on $UED_1$. LAVEA, on the other hand, would try to equalize the number of tasks of $T_1$ and $T_2$ running on the two $UEDs$ thereby resulting in increased interference on $UED_1$ and a high service time. Comparing our schemes with each other, \nameI has inaccuracies in correctly estimating the amount of interference and hence has a higher service time than \nameP. % In the right half of Figure~\ref{fig:running_avg_latency}, \nameI uses online readjustment as described in Section~\ref{subsec:online_readjustment} to correct for the inaccuracies due to the imperfect estimation of the incremental service time matrix. 
For \nameI, looking at the left and right halves of the figure, we see that it starts with a high average service time but slowly converges to the ideal case.
% with a reduction in the average service time from $0.72 s$ to $0.54 s$. 
The online readjustment helps not only in alleviating the inaccuracies because of imperfect information but also handles online heterogeneties like variation in the computational capacity and sporadic availability of $UEDs$. These were not considered in Figure~\ref{fig:running_avg_latency}. We present the evaluation with these heterogeneities involved next.
The incremental service time matrix constructed by \nameI has an average distance of $0.85$ from the true matrix, distance being computed as the Frobenius norm. For a matrix with values in the range $(0.1, 0.6)$, this can be taken to be a medium level of error.

% Note that for the evaluation shown in Figure~\ref{fig:without_online_readjustment}, we have not considered online heterogeneity in terms of variations in the computational capacity of the $UEDs$ or the impact of sporadic availability of $UEDs$. Since our schemes can handle them, we expect our performance to be better than the others in the face of these additional heterogeneties as well. We confirm that in a later evaluation.

% \subsection{Evaluation with Online Readjustment}

\begin{figure}[b]
        \centering
        \begin{subfigure}[h]{0.49\columnwidth}
            \centering
            \includegraphics[width=\columnwidth]{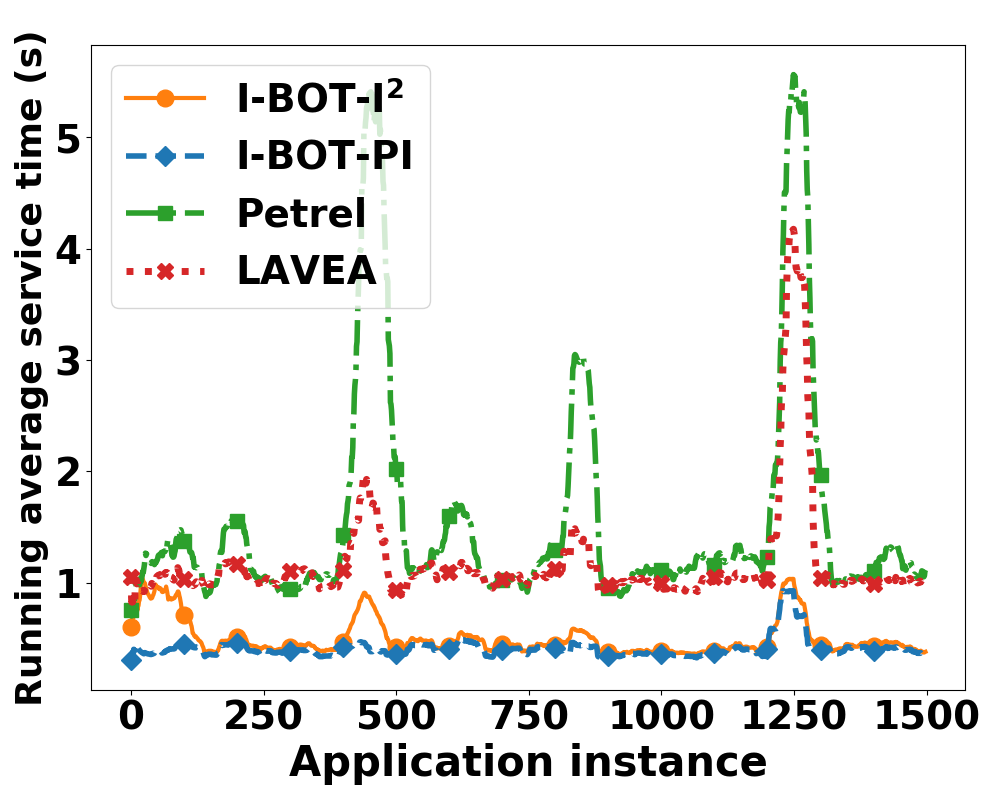}
            \caption[] {Experiment with variation in computation capacity affecting $10\%$ of the tasks}
            \label{fig:10_percent_variation_computation}
        \end{subfigure}
        \hfill
        \begin{subfigure}[h]{0.49\columnwidth}  
            \centering 
            \includegraphics[width=\columnwidth]{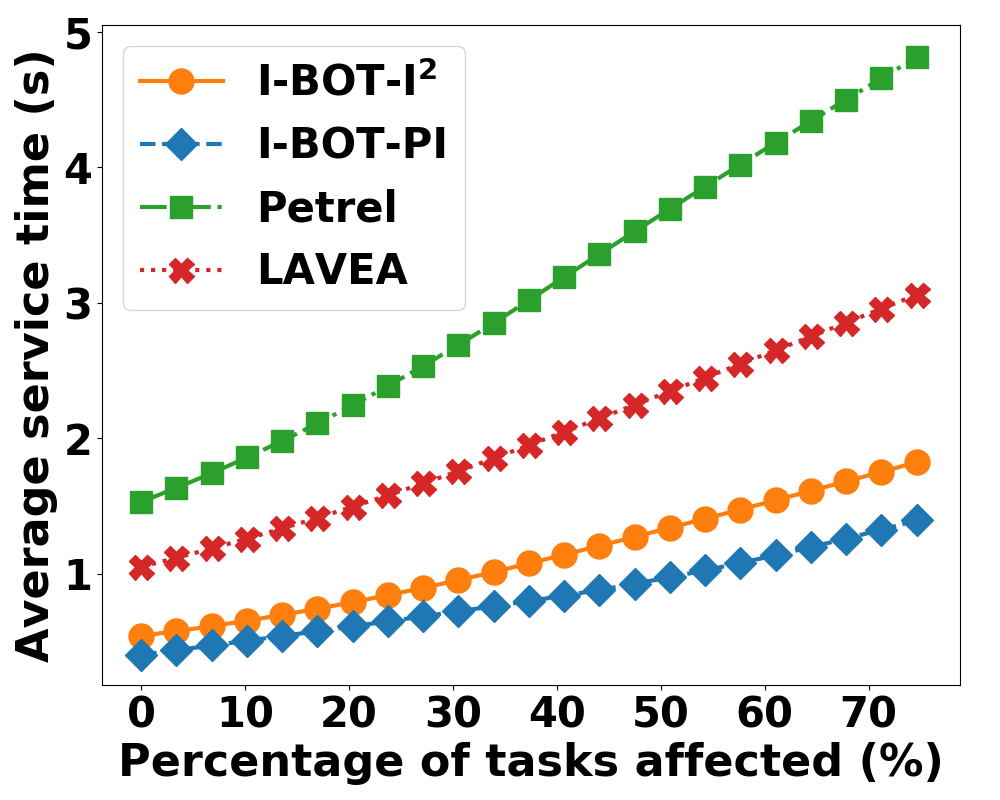}
            \caption[]{Impact of changing variation in the computation capacity}
            \label{fig:changing_percent_variation_computation}
        \end{subfigure}
        \vskip\baselineskip
        \vspace{-15 pt}
        \caption[]
        {Impact of variation in the computation capacity of the $UEDs$ on the service time} 
        \label{fig:variation_in_computation_capacity}
        \vspace{-5 pt}
\end{figure}

% \begin{figure*}[ht]
%         \centering
%         \begin{subfigure}[h]{\columnwidth}
%             \centering
%             \includegraphics[width=\columnwidth, height=5.5cm]{Figures/running_avg_servicetime_lambda_3heterogeneity.png}
%             \caption[] {Impact of variation in computational capacity of the $UEDs$}
%             \label{fig:varying_capacity}
%         \end{subfigure}
%         \hfill
%         \begin{subfigure}[h]{\columnwidth}  
%             \centering 
%             \includegraphics[width=\columnwidth, height=5.5cm]{Figures/devices_unavailable.png}
%             \caption[]{Impact of sporadic availability of $UEDs$}
%             \label{fig:sporadic_availability}
%         \end{subfigure}
%         \vskip\baselineskip
%         \vspace{-15 pt}
%         \caption[]
%         {Comparison of different orchestration schemes with online heterogeneity} 
%         \label{fig:online heterogeneity}
%     \end{figure*}

\subsection{Evaluation with Online Heterogeneity}

% lavea changes from 1.01s to 1.20s = 18.81% 
% Petrel changes from 1.19s to 1.50s = 26.05%
% our perfect info changes from 0.39s to 0.42s = 7.14 
% our imperfect info changes from 0.54s to 0.59s = 9.2%

Online heterogeneity happens due to change in the availability of devices online such as if the owner of a particular $UED$ starts/stops running her personal applications resulting in a change in the available computation capacity of the $UED$ or if a $UED$ enters/exits the system.

\noindent \textbf {Impact of co-located applications on the UEDs:} Our scheme handles the change in computation capacity of $UEDs$ by continuously updating the incremental service time matrix based on the feedback as explained in Section~\ref{subsec:online_readjustment}. Figure~\ref{fig:10_percent_variation_computation} shows the comparison of the running average service time for the orchestration schemes with computation capacity of the $UEDs$ varying for $10\%$ of the scheduled tasks. The spikes in the running average service time occur when the available capacity of one or more $UEDs$ suddenly reduces. It is evident from Figure~\ref{fig:10_percent_variation_computation} that the impact of this variation is the least on our schemes. For other schemes, there is a higher increase in the service time. The average service time increases by $18.81\%$ and $26.05\%$ for LAVEA and Petrel respectively. On the other hand, the increase is only $7.14\%$ and $9.12\%$ for \nameP and \nameI respectively. Also, as the amount of variation in the computation capacity increases, a higher percentage of tasks are affected. With this increase in the percentage of affected tasks, the rate of increase in the average service time is much lower for our schemes compared to the others, as shown in Figure~\ref{fig:changing_percent_variation_computation}. We have not shown comparison with round robin and random schemes here because the impact is significantly higher on those schemes.

\begin{figure}[b]
        \centering
        \begin{subfigure}[h]{0.49\columnwidth}
            \centering
            \includegraphics[width=\columnwidth]{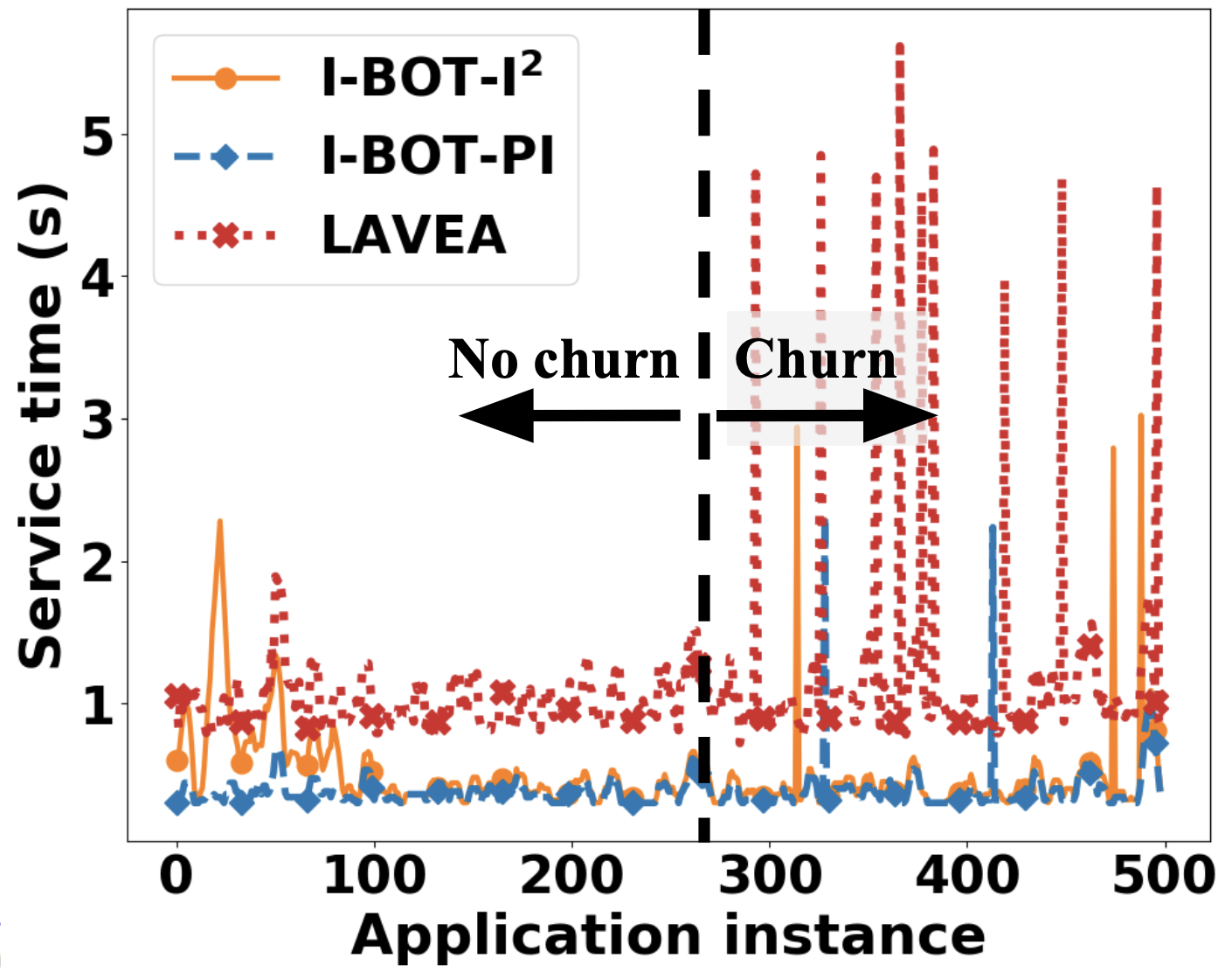}
            \caption[] {Experiment with churn affecting $10\%$ of the tasks}
            \label{fig:10_percent_churn}
        \end{subfigure}
        \hfill
        \begin{subfigure}[h]{0.49\columnwidth}  
            \centering 
            \includegraphics[width=\columnwidth]{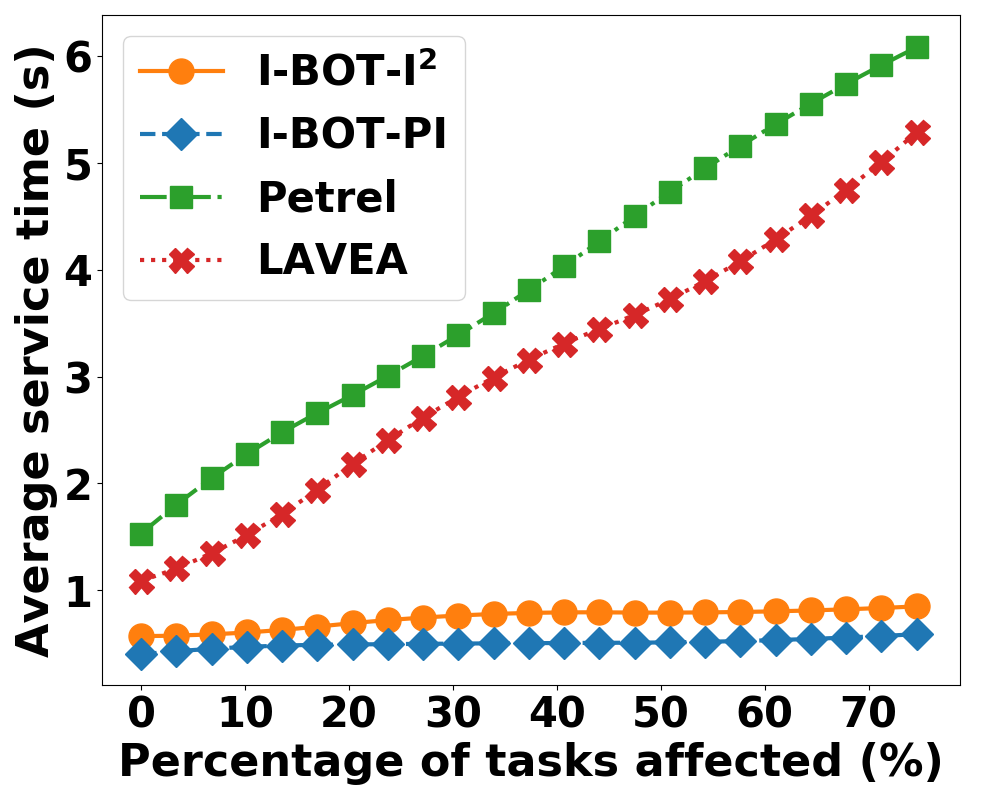}
            \caption[]{Impact of changing amount of churn of the $UEDs$}
            \label{fig:varying_churn}
        \end{subfigure}
        \vskip\baselineskip
        \vspace{-15 pt}
        \caption[]
        {Impact of sporadic availability of the $UEDs$ on the service time} 
        \label{fig:sporadic_availability}
        \vspace{-5 pt}
\end{figure}

\begin{figure*}[t]
\begin{multicols}{3}

     \includegraphics[width=\linewidth]{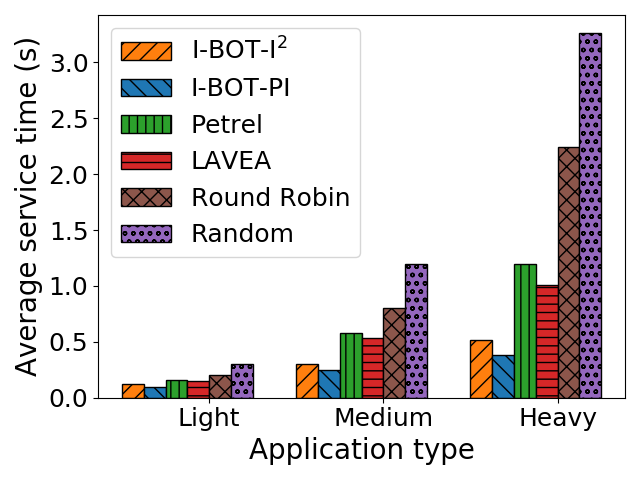}\par\caption{\label{fig:different_application_types}Evaluation with different types of application}
     
    \includegraphics[width=\linewidth, height=3.9cm]{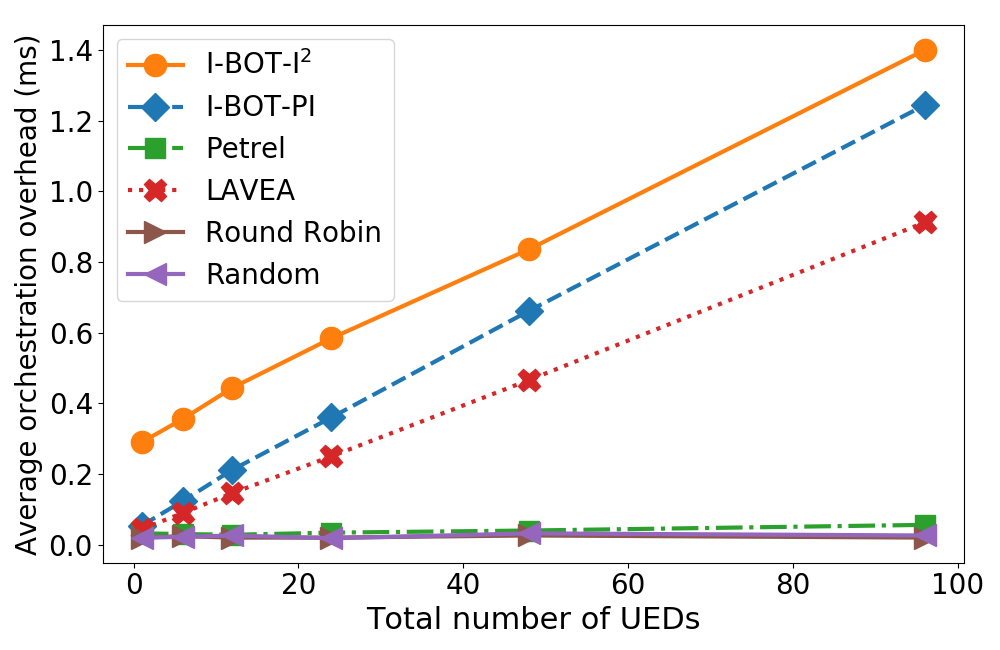}\par\caption{\label{fig:orchestration_overhead}Evaluation of the orchestration overhead}
   
    \includegraphics[width=\linewidth]{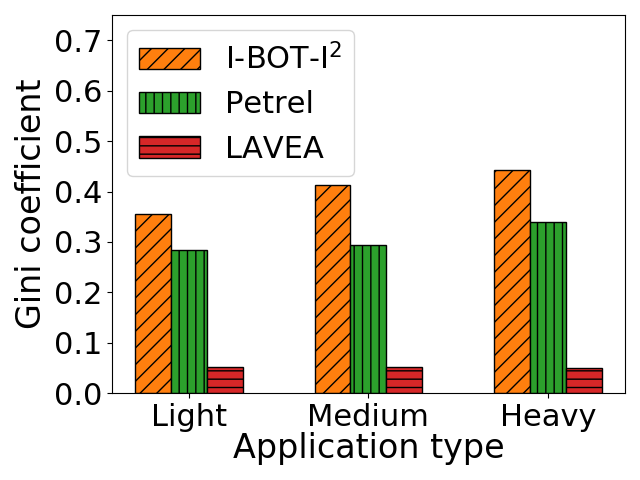} \par\caption{\label{fig:fairness}Evaluation of fairness}
   
\end{multicols}
    \vspace{-15 pt}
\end{figure*}

% lavea changes from 1.01s to 1.27s = 25.74% 
% Petrel changes from 1.19s to 1.56s = 31.09%
% our perfect info changes from 0.39s to 0.41s = 5.12% 
% our imperfect info changes from 0.54s to 0.58s = 7.41%

\noindent \textbf{Impact of churn of UEDs:} In Figure~\ref{fig:sporadic_availability}, we show the impact of sporadic availability of $UEDs$ when one or more $UEDs$ abruptly enter/exit the system. Greater the churn of the $UEDs$, higher would be the percentage of tasks affected. Using the availability history of a $UED$, we predict the probability of the $UED$ being available throughout the task completion. A task is scheduled on a $UED$ only if this probability exceeds the threshold $\gamma$. 
% If a task from an application instance is scheduled on a $UED$ which is unavailable, the task needs to be rescheduled resulting in a higher service time for the instance. 
We have used service time for individual instances instead of a running average in this experiment because it better captures the impact of sporadic availability of $UEDs$. In the left half of Figure~\ref{fig:10_percent_churn} (upto instance 250), all the $UEDs$ are available throughout, whereas in the right half one or more $UEDs$ frequently enter/exit the system resulting in $10\%$ of the tasks being affected due to the churn. The spike in service time occurs when one or more tasks of an application instance are scheduled on a $UED$ which is unavailable or which exits the system before the task completion. Since our schemes predict the availability before task scheduling, the spikes are less frequent and shorter compared to the others. The increase in average service time for our schemes with perfect and imperfect information is $5.12\%$ and $7.41\%$ respectively compared to a significantly higher increase of $25.74\%$ and $31.09\%$ for LAVEA and Petrel respectively. In Figure~\ref{fig:varying_churn}, we show the impact of variation in the churn on the average service time. As the churn increases, a higher percentage of tasks are affected. For our schemes, the average service time increases negligibly with an increase in the churn. However, there is a significant increase in the service time for the two other schemes.

\subsection{Evaluation of Bandwidth Overhead}

\begin{figure}[t]
        \centering
        \begin{subfigure}[h]{0.49\columnwidth}
            \centering
            \includegraphics[width=\columnwidth]{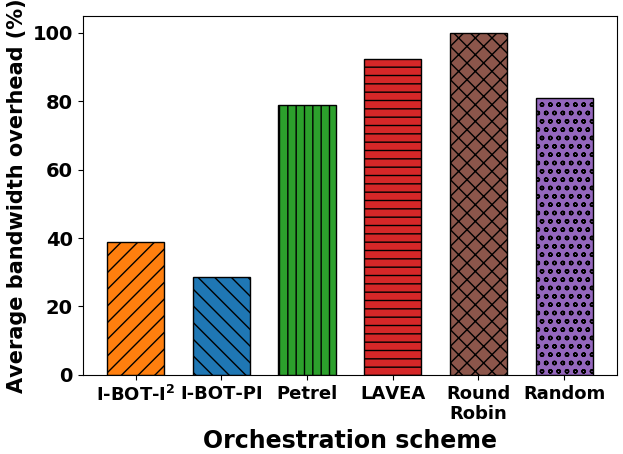}
            \caption[] {Average bandwidth overhead}
            \label{fig:bandwidth_overhead_beta_0.15}
        \end{subfigure}
        \hfill
        \begin{subfigure}[h]{0.49\columnwidth}  
            \centering 
            \includegraphics[width=\columnwidth]{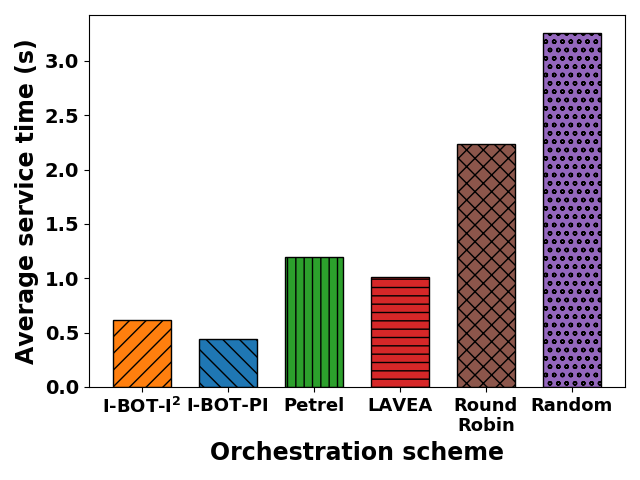}
            \caption[]{Average service time}
            \label{fig:average_service_time_beta_0.15}
        \end{subfigure}
        \vskip\baselineskip
        \vspace{-15 pt}
        \caption[]
        {Comparison of average bandwidth overhead and average service time for the orchestration schemes ($\beta = 0.15$ for our schemes)} 
        \label{fig:bandwidth_overhead}
        \vspace{-15 pt}
\end{figure}

The design of our orchestration scheme uses the parameter $\beta$ that controls the trade off between the average service time and the average bandwidth overhead. In the absence of this design parameter, (i.e., for $\beta = 0$), the average bandwidth overhead for our scheme with perfect and imperfect information is $82\%$ and $85\%$ respectively (100\% means all tasks needing same input data are scheduled on different UEDs). This is comparable to the average bandwidth overhead of the other schemes. However, upon increasing $\beta$, there is a considerable reduction in the average bandwidth overhead of our schemes without a significant increase in the average service time. Figure~\ref{fig:bandwidth_overhead} shows a comparison of the average bandwidth overhead and average service time of the orchestration schemes for $\beta = 0.15$. The average bandwidth overhead for \nameP and \nameI reduces to $28.60\%$ and $30.80\%$ respectively, which is much lower than the other schemes. Meanwhile, the average service times for our two schemes do not increase much and are still lower than the others.

\begin{figure*}[t]
        \centering
        \begin{subfigure}[h]{0.66\columnwidth}
            \centering
            \includegraphics[width=\columnwidth]{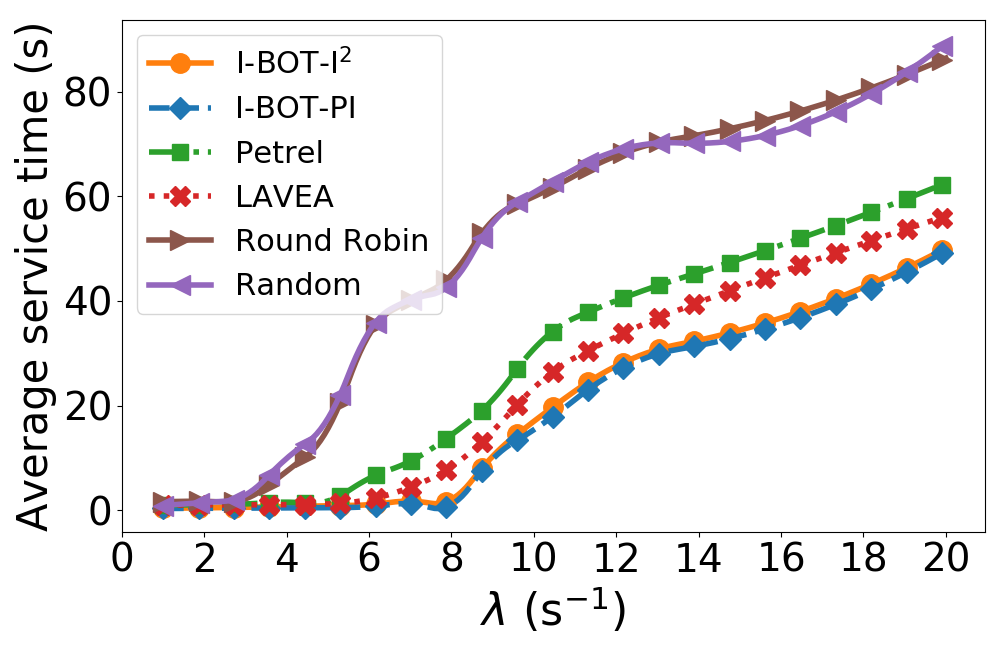}
            \caption[] {Impact of varying arrival rate ($\lambda$) on average service time}
            \label{fig:average_service_time_lambda}
        \end{subfigure}
        \hfill
        \begin{subfigure}[h]{0.66\columnwidth}
            \centering
            \includegraphics[width=\columnwidth]{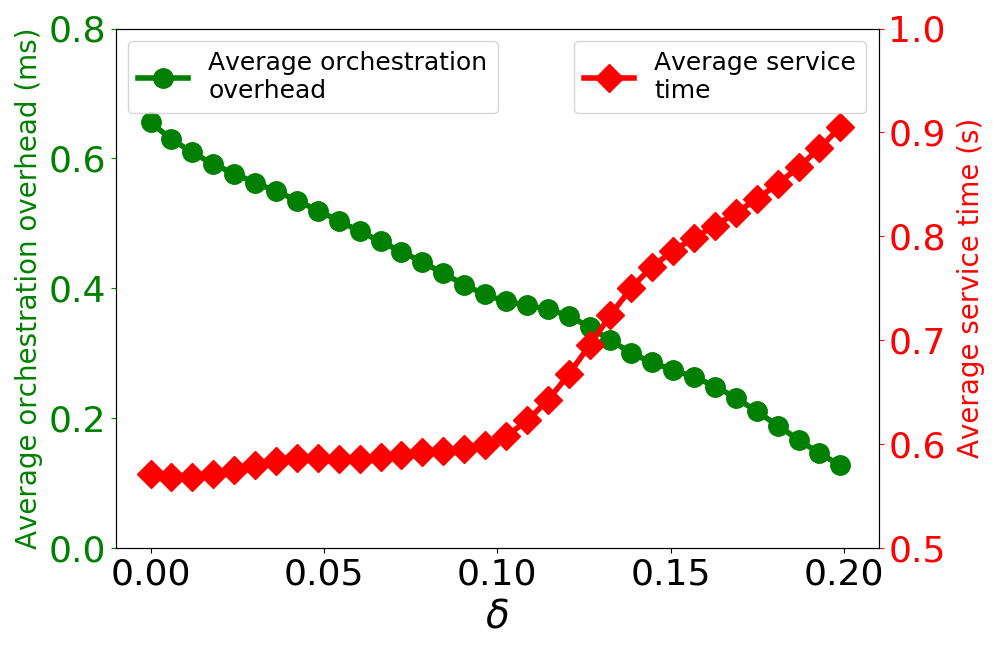}
            \caption[] {Impact of $\delta$ on orchestration overhead and average service time of tasks}
            \label{fig:orchestration_latency_delta}
        \end{subfigure}
        \hfill
        \begin{subfigure}[h]{0.66\columnwidth}  
            \centering 
            \includegraphics[width=\columnwidth]{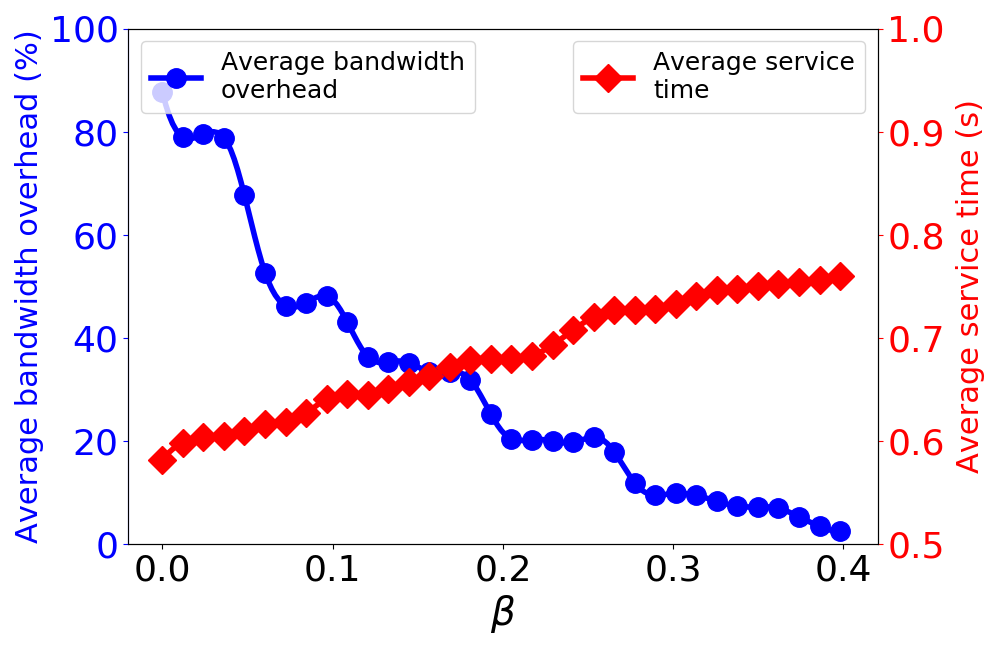}
            \caption[]{Impact of $\beta$ on bandwidth overhead and average service time of tasks}
            \label{fig:bw_latency_beta}
        \end{subfigure}
        \hfill
        \vspace{-5 pt}
        \caption[]
        {Micro evaluations} 
        \label{fig:micro_evaluations}
    \end{figure*}

\subsection{Evaluation with Different Types of Application}
\label{subsec:different_application_types}

Figure~\ref{fig:different_application_types} shows a comparison of the average service time obtained by the orchestration schemes for $500$ instances each of the three different application types: light-weight, medium, and  heavy  (Table~\ref{table:different_applications}). It is evident from Figure~\ref{fig:different_application_types} that there is an advantage in using our orchestration scheme for all the  types of applications. Moreover, this advantage is more pronounced for the heavy application as the tasks involved in such an application have a higher interference with each other and \name schedules the tasks respecting the interference dependencies while the others do not. Most of the latency-sensitive applications that require edge computing belong to this category as they are computationally intensive.

\subsection{Evaluation of Orchestration Overhead}
Here we compare the average orchestration overhead of the schemes as we vary the total number of $UEDs$. For a given number of $UEDs$, as the stream of application instances arrive, we measure the amount of time spent in making the scheduling decisions for every instance and report the average value over $500$ instances. From Figure~\ref{fig:orchestration_overhead}, we can observe that the average orchestration overhead for both our schemes is higher than that for the others. However, it is still negligible given the benefits in terms of the reduction in service time. For instance, the average orchestration overhead of \nameI is only $1.4 ms$ in the presence of $96$ $UEDs$. This accounts for only $0.19\%$ of the average service time of the application instances. Comparing our two variants, the average orchestration overhead is higher for \nameI because of the extra time spent to correct for the inaccuracies due to imperfect information. 

% Note the orchestration overhead does not consider the initial profiling time, which is $1$ minute for the scheme with imperfect information and around $5$ minutes for the scheme with perfect information.

\vspace{-8 pt}

\subsection{Evaluation of Fairness}

In the context of unmanaged edge computing, fairness, defined as balancing the task assignment among multiple $UEDs$, would result in a higher service time because of the substantial heterogeneity among the $UEDs$. We use Gini coefficient to quantify fairness --- a value of $0$ represents perfect equality whereas $1$ represents perfect inequality. Figure~\ref{fig:fairness} shows a comparison of the Gini coefficient for \nameI, Petrel, and LAVEA for the $3$ different application types. As expected, the Gini coefficient is higher for our schemes compared to the others implying a higher inequality in the task distribution among the UEDs. For Petrel, the Gini coefficient is higher than LAVEA, and for round robin, it is equal to $0$ as the tasks are perfectly balanced. We argue that this disparity in task allotment is {\em necessary} for the required improvement in service time because the more powerful UEDs are capable of executing more co-located tasks. 
% Nevertheless, we can be fair in terms of the payment proportional to the usage of the UEDs which was the most popular choice among the survey participants, as shown earlier in Figure~\ref{fig:incentive_model}.

\vspace{-8 pt}

\subsection{Micro Evaluations}

In this section, we evaluate the impact of varying the parameters of our proposed scheme --- application instance arrival rate ($\lambda$), readjustment control parameter ($\delta$), and bandwidth overhead control parameter ($\beta$). The results are presented in Figure~\ref{fig:micro_evaluations}.

% \noindent \textbf{Impact of different percentage of variation in computation capacity of $UEDs$:} Figure~\ref{fig:average_service_time_heterogeneity} shows this evaluation. A $50\%$ variation in computational capacity means that the computational capacity of the $UEDs$ vary $50\%$ of the time. As the percentage of variation increases, the average service time also increases for all the schemes. However, the rate of increase of the average service time is much lower for our schemes compared to the others.

% \noindent \textbf{Impact of varying churn rate of $UEDs$:} Figure~\ref{fig:average_service_time_churn_rate} shows this evaluation. Churn rate is the frequency at which the devices enter/exit the system. A churn rate of $50\%$ indicates that the devices show sporadic availability $50\%$ of the time. Our schemes successfully predict this sporadic behaviour most of the time (> 85\%) and hence, the increase in the average service time is negligible as the churn rate increases. On the other hand, the increase in the service time for the other schemes is significant.

\noindent \textbf{Impact of varying arrival rate ($\lambda$) of application instances:} For a fixed number of $UEDs$, an increase in the arrival rate ($\lambda$) of application instances would ultimately make the system unstable for all the orchestration schemes. However, as shown in Figure~\ref{fig:average_service_time_lambda}, this instability occurs in our schemes for a higher value of $\lambda$ compared to the others, i.e, our schemes can maintain acceptable service time for a higher arrival rate than the other schemes. Also, before the instability sets in, our schemes have lower service time than the others. It is interesting to note that even in the unstable region, the service time for our schemes is lower.

\noindent \textbf{Impact of varying readjustment control parameter ($\delta$):} Figure~\ref{fig:orchestration_latency_delta} shows this evaluation. A low value of $\delta$ implies that gradient descent based online readjustment would be invoked more frequently. Thus there is a higher chance of readjusting to the true matrix A, resulting in a lower average service time. However, more readjustment also means a higher orchestration overhead. 
% As $\delta$ increases, the amount of readjustment goes down, the average service time increases and the orchestration overhead reduces. It is interesting to note that with the increase in 
As $\delta$ increases, the average service time curve is flat in the beginning
% (as $\delta$ goes from $0$ to $0.1$) 
and then increases (for $\delta > 0.1$). This is because while it is true that the amount of readjustment is lower for $\delta = 0.1$ compared to that for $\delta = 0$, it is still sufficient to correct for the online heterogeneities.

\noindent \textbf{Impact of varying bandwidth overhead control parameter ($\beta$):} As $\beta$ increases, the probability of tasks with the same input data being scheduled on the same $UED$ also increases. This results in a reduction in bandwidth overhead as shown in Figure~\ref{fig:bw_latency_beta}. This reduction in the bandwidth overhead comes at the cost of an increase in the service time. However, the increase in the service time is not very significant. For instance, increasing $\beta$ from $0$ to $0.15$ reduces bandwidth overhead by $61\%$ whereas the average service time increases only by $10.3\%$.

\vspace{-8 pt}

\subsection{Theoretical Analysis}
\begin{figure}[b]
\centering
\includegraphics[width=0.85\columnwidth]{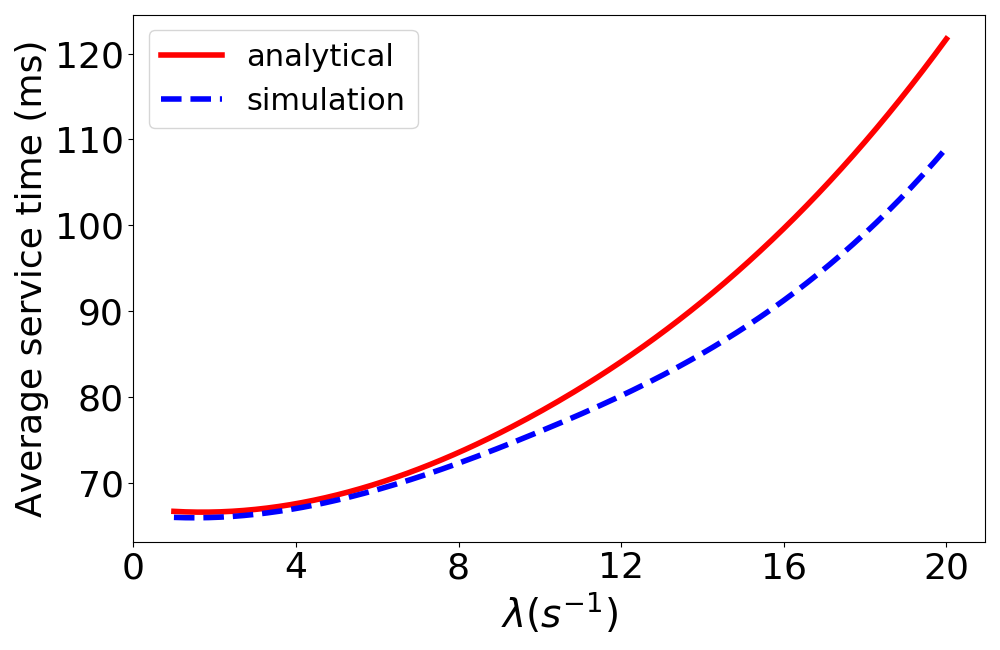}
\vspace{-12pt}
\caption{Comparison of analytical and simulation results}
\vspace{-15pt}
\label{fig:analytical_vs_simulation}
\end{figure}

We present the theoretical analysis of our solution under the following simplifying assumptions. First, we assume that the UEDs are homogeneous 
% have the same inherent capability to process a task, whereas 
and a task of type $k$ has exponentially distributed processing rate $\mu_k$ for $k=[1:N]$, where $\frac{1}{Q} \sum_{k=1}^N \lambda / \mu_k < 1$ for irreducible and stable (\ie positive recurrent) Markov chain. 
Moreover, we assume that tasks of type $1$ to $N$ are dispatched to the chosen UED's queue in order.
Queue state for $UED_q$, $q=[1:Q]$, is then defined by $\phi_n = \{ (0) \} \cup \{(t_1,t_2,\ldots,t_n) \vert n \ge 1 \}$, where $t_i$ is the type of the $i$th task in the (type independent) FIFO order ($t_1$ is the type of a task being served) and $(0)$ represents the empty system. 
The processed rate is then determined by the type of a task being served and uniquely determined by queue length $i$ as follows: $\ceil*{\frac{i}{N}} N - i + 1$.
Due to space limitation, we present the results without detailed explanations. For the proof, the interested reader is referred to the anonymized supplemental material~\cite{supplementary-material}.
\begin{lemma} \label{lemma:transition_rates}
    Under our proposed solution, the transition rates $q_{i,j}({\boldsymbol \pi})$ given distribution ${\boldsymbol \pi}$ for $j \neq i$ is given by 
    \[
    q_{i,j}({\boldsymbol \pi}) = 
        \left\{
            \begin{array}{ll}
                \mu_{\ceil*{\frac{i}{N}} N - i + 1} & \mbox{if $j=i-1$}, \\
                \frac{1 - (Q-1) \sum_{l=0}^{i-1} \pi_l}{1 + (Q-1) \sum_{l=0}^i \pi_l} & \mbox{if $j=i+N$, $i < \tau_{\boldsymbol \pi}$}, \\
                0, & \mbox{otherwise,}
            \end{array}
        \right.
    \]
    where $\tau_{\boldsymbol \pi} = \min \{ j : \sum_{l=0}^{j-1} \pi_l \ge \frac{1}{Q-1} \}$ and $\pi_l$ denotes the stationary distribution of UED queue, i.e., the probability that the queue size is $l$ at a UED.
\end{lemma}
Intuitively, $\tau_{\boldsymbol \pi}$ indicates the queue length so that the probability that a UED with queue size $i (\ge \tau_{\boldsymbol \pi})$ receives $N$ tasks is $0$.
Based on Lemma \ref{lemma:transition_rates}, we can calculate the stationary distribution of the queue length of a single UED numerically by finding $\hat{\pi}$ that satisfies the global balance equation.
The expected service time $T_Q(\lambda,\mu_1,\cdots,\mu_N)$ of an application instance that is dispatched to $Q$ UEDs is then
\begin{align*}
    \sum_{r=0}^{N-1} \left[ \sum_{i=1}^{\infty} \left( \floor*{\frac{i-1}{N}} \sum_{l=1}^N \frac{1}{\mu_l} + \ind_{\ceil*{\frac{i}{N}} N - i + 1 - r \ge 1} \sum_{m = \ceil*{\frac{i}{N}} N - i + 1 - r}^{N-r} \frac{1}{\mu_m} + \right. \right. \\
    \left. \left. \ind_{\ceil*{\frac{i}{N}} N - i + 1 - r < 1} \sum_{m = N-r}^{\ceil*{\frac{i}{N}} N - i + 1 - r + N} \frac{1}{\mu_m} \right)
    \cdot \left\{ \left( \sum_{j=i-1}^{\infty} \pi_j \right)^Q - \left( \sum_{j=i}^{\infty} \pi_j \right)^Q \right\} \right].
\end{align*}
% The expected service time of an application instance that is dispatched to UEDs is then
% \begin{align*}
%     T_Q(\lambda,\mu_1,\cdots,\mu_N) = N \sum_{i=1}^{\infty} \left( \floor*{\frac{i-1}{N}} \sum_{l=1}^N \frac{1}{\mu_l} + \sum_{m = \ceil*{\frac{i}{N}} N - i + 1}^N \frac{1}{\mu_m} \right) \\ 
%     \cdot \left\{ \left( \sum_{j=i-1}^{\infty} \pi_j \right)^Q - \left( \sum_{j=i}^{\infty} \pi_j \right)^Q \right\}.
% \end{align*}
We compare the expected service times from analysis and simulations, as shown in Figure \ref{fig:analytical_vs_simulation}, where $N=2,~Q=3,~\lambda=[1:20],~\mu_1=10$, and $\mu_2=30$.
It demonstrates that the analytical result serves as a worst case upper bound  for the service time as it assumes serial processing, while in reality multiple tasks can be concurrently processed. The worst case will occur in practice if each task is intensive enough to occupy the entire UED. The divergence between analytical and simulation results increases as the load increases, in which case the simulation allows for more parallel processing. 

%% file: sec_discussion.tex
\section{Discussion}
\label{sec:discussion}
In this section, we present extensions of \name needed to handle some important cases. {\em First}, we have only considered one form of dependency among the tasks --- same input data requirements for certain tasks. In practice, however, the tasks can also have control  dependency, which is often represented by a directed acyclic graph (DAG) of tasks. Our algorithm can conceptually be easily combined with scheduling algorithms that operate over DAGs; there is significant work in the context of conventional computing systems~\cite{sakellariou2004hybrid, bosilca2012dague} and some emerging work in the context of edge computing systems~\cite{khare2019linearize}. {\em Second}, the linearity in the task interference plots may not hold if the number of tasks running on a UED is large enough to cause a discontinuous change. This is a common occurrence with mapping of resource availability to performance metrics (such as, latency)~\cite{falsafi2014primer}, say if the working set of the program spills over from one level of cache into a lower (and higher latency) level of cache. In this case, in \name, a higher-order characterization of the interference plots would be needed (say, quadratic or piece-wise linear) and failure of tasks must also be considered. Next, for simplicity, we fix the order in which the tasks belonging to an application instance are considered for offloading, namely, the same order in which they are enumerated in the application description. This is akin to greedy scheduling with respect to task order and a more optimal scheduling can happen if we use non-local information, such as, through dynamic programming. 
% This is because exhaustively considering all possible permutations increases the orchestration overhead without significantly reducing the service time. However, a detailed evaluation of this trade-off should be conducted to explore scope of further improvements in our scheme. Finally, while we use gradient descent for online readjustment, other optimization schemes like genetic algorithm, Levenberg–Marquardt algorithm, etc. can be considered. 

%% file: sec_related_work.tex
\section{Related Work}
\label{sec:related_work}

In this section we contrast our work with the other efforts in the field of task scheduling in heterogeneous edge computing systems.

\noindent \textbf{Low latency edge scheduling:} 
Petrel~\cite{petrel} and LAVEA~\cite{lavea} propose orchestration schemes aimed at minimizing the service time in a multi-edge collaborative environment. We have shown that \name outperforms these schemes in terms of the service time and bandwidth overhead in a heterogeneous unmanaged edge computing setting. MSGA~\cite{MSGA} jointly studies the task and network flow scheduling and uses a multistage greedy algorithm to minimize the completion time of the application. 
% COOPER-SCHED~\cite{Cooper_sched} proposes an extended marriage algorithm for a cooperative game based job mapping on multiple heterogeneous servers to maximize the number of jobs whose deadlines are satisfied. 
In~\cite{Tseng2018}, a gateway-based edge computing service model has been proposed to reduce the latency of data transmission and the network bandwidth. Low latency task scheduling schemes for edge have also been proposed in ~\cite{Wang2020,low_latency1,low_latency2}. However, all of these works are in the context of managed edge and do not consider the unique challenges introduced by unmanaged edge, such as the lack of monitoring information, heterogeneity, and unexpected entry-exits. One exception to this is CoGTA~\cite{zhang2018cooperative}, which considers scheduling of delay-sensitive social sensing tasks on a heterogeneous unmanaged edge. However, its main focus is on devices that are not trusted and therefore it formulates a game-theoretic technique to perform the task allocation. Its performance in a benign setting like ours is likely to be sub-optimal. 

\noindent \textbf{Availability and Interference based edge scheduling:} There have been a few efforts that take into account the availability and interference while devising strategies for task scheduling on the edge. An overhead-optimizing task scheduling strategy has been proposed in~\cite{ad_hoc_1} for ad-hoc based edge computing nodes formed by a group of mobile devices.~\cite{Aral2019} proposes a score based edge service scheduling algorithm that evaluates network, compute, and reliability capabilities of edge nodes. However, these works rely on sharing monitoring information which can be a huge overhead in highly dynamic environments. Also, the time and energy consumption models are theoretical and have not been tested on real systems. INDICES~\cite{INDICES} proposes a performance-aware scheme for migrating services from cloud to edge while taking into account the interference caused by co-located applications. However, this is geared towards service migration and not task scheduling. Also, it does not consider the impact of online variations in the availability and compute capabilities of edge devices.

\noindent \textbf{Energy efficient edge scheduling:} A lot of existing works~\cite{dvfs_1,dvfs_2,dvfs_3,dvfs_4} utilize dynamic voltage-frequency scaling (DVFS), which is an attractive method for reducing energy consumption in heterogeneous computing systems.
% SB (7/10/20): Not as relevant. 
% Hybrid DVFS schemes that employ Deep Learning to combine multiple DVFS technologies have been proposed in~\cite{hybrid_dvfs_1,hybrid_dvfs_2}. 
ESTS~\cite{ESTS} deals with the problem of scheduling a group of tasks, optimizing both the schedule length and energy consumption. They formulate the problem as a joint linear programming problem and propose a heuristic algorithm to solve it. In~\cite{mec_offloading}, a computational offloading framework has been proposed which minimizes the total energy consumption and execution latency by coupling task allocation decisions and frequency scaling. The paper~\cite{zhang2017energy} also performs joint optimization of energy and latency through a rigorously formulated and solved  mixed integer  nonlinear  problem  (MINLP)  for  computation  offloading and resource allocation. However, the execution models used in these works 
% employ CPU frequency for latency estimation which 
do not consider the impact of online heterogeneties in the computation capacity or the effect of interference.
% on the execution time and is not suitable for an unmanaged scenario.

\noindent \textbf{Volunteer or opportunistic computing:} In a completely different context, under the moniker "volunteer computing", a slew of works designed solutions to utilize under-utilized compute nodes (such as, on a university campus) or mobile devices to run large-scale parallel applications. An example of the former is HTCondor~\cite{epema1996worldwide} and an example of the latter is Femtocloud~\cite{habak2015femto}. Our design borrows some features from Femtocloud (identifying devices with spare capacity and some stability); however, Femtocloud did not have to deal with the majority of the challenges that we solve here (great heterogeneity from a compute, network, and application standpoint, unknown tasks, runtime variations due to interference).

%% file: sec_conclusion.tex
\vspace{-5pt}
\section{Conclusion}
\label{sec:conclusion}

In this paper, we presented a novel Interference Based Orchestration of Tasks (\name) for unmanaged edge computing that utilizes personal devices as edge nodes for task execution. We identified three new challenges in orchestrating application tasks in the unmanaged edge scenario, due to which prior edge schedulers fail --- heterogeneity in devices, runtime variation in available compute capacity, and sporadic availability of devices. We introduced three design innovations in \name to handle these challenges and thus minimize the service time and bandwidth overhead of latency-sensitive applications.  We extensively evaluated our system using real-world experiments and simulations. Results show that compared to existing approaches (two intuitive baselines and two state-of-the-art ones, LAVEA and Petrel), \name significantly reduces average service time and bandwidth overhead of applications by at least $61\%$ and $56\%$ respectively. We also demonstrated that in the presence of online variability, which is an inherent characteristic of unmanaged systems, the reduction in service time and bandwidth overhead due to \name is more prominent.

%% file: main.bbl
\begin{thebibliography}{10}

\bibitem{lambda_edge}
Amazon: Lambda@edge.
\newblock \url{https://aws.amazon.com/lambda/edge/}, 2020.
\newblock Accessed: 2020-06-28.

\bibitem{cisco_edge}
Cisco: Establishing the edge.
\newblock
  \url{https://www.cisco.com/c/en/us/solutions/service-provider/edge-computing/establishing-the-edge.html},
  2020.
\newblock Accessed: 2020-06-28.

\bibitem{google_edge}
Google: Edge network.
\newblock \url{https://peering.google.com}, 2020.
\newblock Accessed: 2020-06-28.

\bibitem{supplementary-material}
\name: Anonymized supplementary material.
\newblock \url{https://bit.ly/3iTuYaT}, 2020.
\newblock Accessed: 2020-07-10.

\bibitem{fog_2}
{\sc {Aazam}, M., and {Huh}, E.}
\newblock Fog computing and smart gateway based communication for cloud of
  things.
\newblock In {\em 2014 International Conference on Future Internet of Things
  and Cloud\/} (2014), pp.~464--470.

\bibitem{mdc_1}
{\sc {Aazam}, M., and {Huh}, E.}
\newblock Fog computing micro datacenter based dynamic resource estimation and
  pricing model for iot.
\newblock In {\em 2015 IEEE 29th International Conference on Advanced
  Information Networking and Applications\/} (2015), pp.~687--694.

\bibitem{Aral2019}
{\sc Aral, A., Brandic, I., Uriarte, R.~B., De~Nicola, R., and Scoca, V.}
\newblock Addressing application latency requirements through edge scheduling.
\newblock {\em Journal of Grid Computing 17}, 4 (Dec 2019), 677--698.

\bibitem{fog_1}
{\sc Bonomi, F., Milito, R., Zhu, J., and Addepalli, S.}
\newblock Fog computing and its role in the internet of things.
\newblock In {\em Proceedings of the First Edition of the MCC Workshop on
  Mobile Cloud Computing\/} (New York, NY, USA, 2012), MCC ’12, Association
  for Computing Machinery, p.~13–16.

\bibitem{bosilca2012dague}
{\sc Bosilca, G., Bouteiller, A., Danalis, A., Herault, T., Lemarinier, P., and
  Dongarra, J.}
\newblock Dague: A generic distributed dag engine for high performance
  computing.
\newblock {\em Parallel Computing 38}, 1-2 (2012), 37--51.

\bibitem{paragon}
{\sc Delimitrou, C., and Kozyrakis, C.}
\newblock Paragon: Qos-aware scheduling for heterogeneous datacenters.
\newblock In {\em Proceedings of the Eighteenth International Conference on
  Architectural Support for Programming Languages and Operating Systems\/} (New
  York, NY, USA, 2013), ASPLOS ’13, Association for Computing Machinery,
  p.~77–88.

\bibitem{mec_offloading}
{\sc {Dinh}, T.~Q., {Tang}, J., {La}, Q.~D., and {Quek}, T. Q.~S.}
\newblock Offloading in mobile edge computing: Task allocation and
  computational frequency scaling.
\newblock {\em IEEE Transactions on Communications 65}, 8 (2017), 3571--3584.

\bibitem{epema1996worldwide}
{\sc Epema, D.~H., Livny, M., van Dantzig, R., Evers, X., and Pruyne, J.}
\newblock A worldwide flock of condors: Load sharing among workstation
  clusters.
\newblock {\em Future Generation Computer Systems 12}, 1 (1996), 53--65.

\bibitem{falsafi2014primer}
{\sc Falsafi, B., and Wenisch, T.~F.}
\newblock A primer on hardware prefetching.
\newblock {\em Synthesis Lectures on Computer Architecture 9}, 1 (2014), 1--67.

\bibitem{autonomous_cars}
{\sc {Fridman}, L., {Brown}, D.~E., {Glazer}, M., {Angell}, W., {Dodd}, S.,
  {Jenik}, B., {Terwilliger}, J., {Patsekin}, A., {Kindelsberger}, J., {Ding},
  L., {Seaman}, S., {Mehler}, A., {Sipperley}, A., {Pettinato}, A., {Seppelt},
  B.~D., {Angell}, L., {Mehler}, B., and {Reimer}, B.}
\newblock Mit advanced vehicle technology study: Large-scale naturalistic
  driving study of driver behavior and interaction with automation.
\newblock {\em IEEE Access 7\/} (2019), 102021--102038.

\bibitem{mdc_2}
{\sc Greenberg, A., Hamilton, J., Maltz, D.~A., and Patel, P.}
\newblock The cost of a cloud: Research problems in data center networks.
\newblock {\em SIGCOMM Comput. Commun. Rev. 39}, 1 (Dec. 2009), 68–73.

\bibitem{habak2015femto}
{\sc Habak, K., Ammar, M., Harras, K.~A., and Zegura, E.}
\newblock Femto clouds: Leveraging mobile devices to provide cloud service at
  the edge.
\newblock In {\em 2015 IEEE 8th international conference on cloud computing\/}
  (2015), IEEE, pp.~9--16.

\bibitem{low_latency1}
{\sc {Han}, J., and {Wang}, D.}
\newblock Edge scheduling algorithms in parallel and distributed systems.
\newblock In {\em 2006 International Conference on Parallel Processing
  (ICPP'06)\/} (2006), pp.~147--154.

\bibitem{low_latency2}
{\sc {He}, T., {Khamfroush}, H., {Wang}, S., {La Porta}, T., and {Stein}, S.}
\newblock It's hard to share: Joint service placement and request scheduling in
  edge clouds with sharable and non-sharable resources.
\newblock In {\em 2018 IEEE 38th International Conference on Distributed
  Computing Systems (ICDCS)\/} (2018), pp.~365--375.

\bibitem{khare2019linearize}
{\sc Khare, S., Sun, H., Gascon-Samson, J., Zhang, K., Gokhale, A., Barve, Y.,
  Bhattacharjee, A., and Koutsoukos, X.}
\newblock Linearize, predict and place: minimizing the makespan for edge-based
  stream processing of directed acyclic graphs.
\newblock In {\em Proceedings of the 4th ACM/IEEE Symposium on Edge
  Computing\/} (2019), pp.~1--14.

\bibitem{dvfs_4}
{\sc {Kimura}, H., {Sato}, M., {Hotta}, Y., {Boku}, T., and {Takahashi}, D.}
\newblock Emprical study on reducing energy of parallel programs using slack
  reclamation by dvfs in a power-scalable high performance cluster.
\newblock In {\em 2006 IEEE International Conference on Cluster Computing\/}
  (2006), pp.~1--10.

\bibitem{dvfs_2}
{\sc {Li}, D., and {Wu}, J.}
\newblock Energy-aware scheduling for frame-based tasks on heterogeneous
  multiprocessor platforms.
\newblock In {\em 2012 41st International Conference on Parallel Processing\/}
  (2012), pp.~430--439.

\bibitem{ESTS}
{\sc {Li}, K., {Tang}, X., and {Li}, K.}
\newblock Energy-efficient stochastic task scheduling on heterogeneous
  computing systems.
\newblock {\em IEEE Transactions on Parallel and Distributed Systems 25}, 11
  (2014), 2867--2876.

\bibitem{petrel}
{\sc {Lin}, L., {Li}, P., {Xiong}, J., and {Lin}, M.}
\newblock Distributed and application-aware task scheduling in edge-clouds.
\newblock In {\em 2018 14th International Conference on Mobile Ad-Hoc and
  Sensor Networks (MSN)\/} (2018), pp.~165--170.

\bibitem{power_of_two}
{\sc {Mitzenmacher}, M.}
\newblock The power of two choices in randomized load balancing.
\newblock {\em IEEE Transactions on Parallel and Distributed Systems 12}, 10
  (2001), 1094--1104.

\bibitem{GA}
{\sc {Page}, A.~J., and {Naughton}, T.~J.}
\newblock Dynamic task scheduling using genetic algorithms for heterogeneous
  distributed computing.
\newblock In {\em 19th IEEE International Parallel and Distributed Processing
  Symposium\/} (2005), pp.~8 pp.--.

\bibitem{panta2011efficient}
{\sc Panta, R.~K., Bagchi, S., and Midkiff, S.~P.}
\newblock Efficient incremental code update for sensor networks.
\newblock {\em ACM Transactions on Sensor Networks (TOSN) 7}, 4 (2011), 1--32.

\bibitem{dvfs_3}
{\sc Rizvandi, N.~B., Taheri, J., and Zomaya, A.~Y.}
\newblock Some observations on optimal frequency selection in dvfs-based energy
  consumption minimization.
\newblock {\em Journal of Parallel and Distributed Computing 71}, 8 (2011),
  1154 -- 1164.

\bibitem{MSGA}
{\sc {Sahni}, Y., {Cao}, J., and {Yang}, L.}
\newblock Data-aware task allocation for achieving low latency in collaborative
  edge computing.
\newblock {\em IEEE Internet of Things Journal 6}, 2 (2019), 3512--3524.

\bibitem{sakellariou2004hybrid}
{\sc Sakellariou, R., and Zhao, H.}
\newblock A hybrid heuristic for dag scheduling on heterogeneous systems.
\newblock In {\em 18th International Parallel and Distributed Processing
  Symposium\/} (2004), IEEE, p.~111.

\bibitem{edge_computing}
{\sc Satyanarayanan, M.}
\newblock The emergence of edge computing.
\newblock {\em Computer 50}, 1 (Jan 2017), 30--39.

\bibitem{cloudlets_1}
{\sc Satyanarayanan, M., Bahl, P., Caceres, R., and Davies, N.}
\newblock The case for vm-based cloudlets in mobile computing.
\newblock {\em IEEE Pervasive Computing 8}, 4 (Oct. 2009), 14--23.

\bibitem{fault_tolerance_1}
{\sc Schlichting, R.~D., and Schneider, F.~B.}
\newblock Fail-stop processors: An approach to designing fault-tolerant
  computing systems.
\newblock {\em ACM Trans. Comput. Syst. 1}, 3 (Aug. 1983), 222–238.

\bibitem{fault_tolerance_2}
{\sc {Schneider}, F.~B., and {Lidong Zhou}}.
\newblock Implementing trustworthy services using replicated state machines.
\newblock {\em IEEE Security \& Privacy 3}, 5 (2005), 34--43.

\bibitem{INDICES}
{\sc {Shekhar}, S., {Chhokra}, A.~D., {Bhattacharjee}, A., {Aupy}, G., and
  {Gokhale}, A.}
\newblock Indices: Exploiting edge resources for performance-aware cloud-hosted
  services.
\newblock In {\em 2017 IEEE 1st International Conference on Fog and Edge
  Computing (ICFEC)\/} (2017), pp.~75--80.

\bibitem{ad_hoc_1}
{\sc {Tianze}, L., {Muqing}, W., {Min}, Z., and {Wenxing}, L.}
\newblock An overhead-optimizing task scheduling strategy for ad-hoc based
  mobile edge computing.
\newblock {\em IEEE Access 5\/} (2017), 5609--5622.

\bibitem{Tseng2018}
{\sc Tseng, C.-W., Tseng, F.-H., Yang, Y.-T., Liu, C.-C., and Chou, L.-D.}
\newblock Task scheduling for edge computing with agile vnfs on-demand service
  model toward 5g and beyond.
\newblock {\em Wireless Communications and Mobile Computing 2018\/} (Jul 2018),
  7802797.

\bibitem{cloudlets_2}
{\sc Verbelen, T., Simoens, P., De~Turck, F., and Dhoedt, B.}
\newblock Cloudlets: bringing the cloud to the mobile user.
\newblock In {\em 3rd ACM Workshop on Mobile Cloud Computing and Services,
  Proceedings\/} (2012), Ghent University, Department of Information
  technology, pp.~29--35.

\bibitem{Wang2020}
{\sc Wang, S., Li, Y., Pang, S., Lu, Q., Wang, S., and Zhao, J.}
\newblock A task scheduling strategy in edge-cloud collaborative scenario based
  on deadline.
\newblock {\em Scientific Programming 2020\/} (Mar 2020), 3967847.

\bibitem{availability_prediction}
{\sc {Xiaojuan Ren}, {Seyong Lee}, {Eigenmann}, R., and {Bagchi}, S.}
\newblock Resource availability prediction in fine-grained cycle sharing
  systems.
\newblock In {\em 2006 15th IEEE International Conference on High Performance
  Distributed Computing\/} (2006), pp.~93--104.

\bibitem{zenith}
{\sc {Xu}, J., {Palanisamy}, B., {Ludwig}, H., and {Wang}, Q.}
\newblock Zenith: Utility-aware resource allocation for edge computing.
\newblock In {\em 2017 IEEE International Conference on Edge Computing
  (EDGE)\/} (2017), pp.~47--54.

\bibitem{lavea}
{\sc Yi, S., Hao, Z., Zhang, Q., Zhang, Q., Shi, W., and Li, Q.}
\newblock Lavea: Latency-aware video analytics on edge computing platform.
\newblock In {\em Proceedings of the Second ACM/IEEE Symposium on Edge
  Computing\/} (New York, NY, USA, 2017), SEC ’17, Association for Computing
  Machinery.

\bibitem{zhang2018cooperative}
{\sc Zhang, D., Ma, Y., Zheng, C., Zhang, Y., Hu, X.~S., and Wang, D.}
\newblock Cooperative-competitive task allocation in edge computing for
  delay-sensitive social sensing.
\newblock In {\em 2018 IEEE/ACM Symposium on Edge Computing (SEC)\/} (2018),
  IEEE, pp.~243--259.

\bibitem{zhang2017energy}
{\sc Zhang, J., Hu, X., Ning, Z., Ngai, E. C.-H., Zhou, L., Wei, J., Cheng, J.,
  and Hu, B.}
\newblock Energy-latency tradeoff for energy-aware offloading in mobile edge
  computing networks.
\newblock {\em IEEE Internet of Things Journal 5}, 4 (2017), 2633--2645.

\bibitem{dvfs_1}
{\sc {Zhuravlev}, S., {Saez}, J.~C., {Blagodurov}, S., {Fedorova}, A., and
  {Prieto}, M.}
\newblock Survey of energy-cognizant scheduling techniques.
\newblock {\em IEEE Transactions on Parallel and Distributed Systems 24}, 7
  (2013), 1447--1464.

\end{thebibliography}
